\documentclass[prb,twocolumn,aps,showpacs,superscriptaddress,floatfix] {revtex4}
\usepackage[dvips]{graphicx}
\usepackage{dcolumn}
\usepackage{bm}
\usepackage{latexsym}
\usepackage{amsmath}
\usepackage{amssymb}
\usepackage{multirow}
\usepackage{subfigure}
\usepackage[colorlinks,bookmarks=false,citecolor=blue,linkcolor=red,urlcolor=blue]{hyperref}

\def\id{\,\mbox{l\hspace{-0.55em}1}}

\begin{document}

\title{Phase diagram of the fully frustrated transverse-field
Ising model  \\ on the honeycomb lattice}

\author{T. Coletta}
\affiliation{Institute of Theoretical Physics, EPF Lausanne, CH-1015 Lausanne, Switzerland}
\author{J.-D. Picon}
\affiliation{Institute of Theoretical Physics, EPF Lausanne, CH-1015 Lausanne, Switzerland}
\author{S.~E. Korshunov}
\affiliation{L. D. Landau Institute for Theoretical Physics,
142432 Chernogolovka, Russia}
\author{F. Mila}
\affiliation{Institute of Theoretical Physics, EPF Lausanne, CH-1015 Lausanne, Switzerland}
\date{\today}


\begin{abstract}
Motivated  by the current interest in the quantum dimer model on the
triangular lattice, we investigate the phase diagram of the closely
related fully-frustrated transverse field Ising model on the honeycomb
lattice using classical and semi-classical approximations. We show that,
in addition to the fully polarized phase at large field, the classical
model possesses a multitude of phases that break the translational symmetry which,
in the dimer language, correspond to a plaquette phase and a columnar
phase separated by an infinite cascade of mixed phases.
The modification of the phase diagram by quantum fluctuations
has been investigated in the context of linear spin-wave theory.
The extrapolation of the semiclassical energies suggests that
the plaquette phase extends down to
zero field for spin 1/2, in agreement with the $\sqrt{12}\times\sqrt{12}$
phase of the quantum dimer model on the triangular lattice with only
kinetic energy.
\end{abstract}

\maketitle

%**************************************************************************
%**************************************************************************

\section{{Introduction}}

Quantum dimer models have emerged as one of the main paradigms in the
investigation of quantum spin liquids. The Rokhsar-Kivelson (RK) quantum
dimer model (QDM), which includes a potential interaction of amplitude $V$
between dimers facing each other and a kinetic term of amplitude $t$
flipping them around rhombic plaquettes, has recently attracted special
attention. The main reason comes from the presence on the
triangular lattice of a resonating valence bond (RVB) phase first
discovered by Moessner and Sondhi\cite{moessner2} and extensively studied
since then using zero temperature Green's function quantum Monte Carlo
(GFQMC).\cite{ralko1,ralko2,ralko3} Exact results have been obtained at
the RK point ($V/t=1$), where the sum of all configurations can be proven
to be a ground state, \cite{moessner2} and at $V>t$, where the
non-flippable configurations are the ground states. Analytical results
have also been obtained in the limit $V/t\rightarrow -\infty$, where
columnar states have been shown to be selected. However, in the
intermediate range below the RK point, most of what is known about the
model is based on numerical simulations.

A closely related model for which a number of analytical results have
already been obtained is the fully frustrated transverse field Ising
model (FFTFIM) on the honeycomb lattice defined by the Hamiltonian:
\begin{equation} \label{eq:Hamiltonian of the model}
 H=-\frac{J}{S^2}\sum_{\left\langle i,j\right\rangle}{M_{ij}S_i^zS_j^z}
 -\frac{\Gamma}{S}\sum_{i}{S_i^x}\,,
\end{equation}
where $\Gamma>0$ is the transverse magnetic field, $J>0$ is the
coupling constant of the Ising interaction term, $\left\langle
i,j\right\rangle$ denotes pairs of nearest neighbors on the honeycomb
lattice, and $M_{ij}=\pm 1$ is such that for each hexagon of the lattice
the number of antiferromagnetic bonds ($M_{ij}=-1$) is odd, different
choices of $M_{ij}$ corresponding to the same model up to the rotation of
some spins by $\pi$ around the $x$ axis\cite{Villain}. Transverse field Ising models
have been the subject of intense investigations over the years.
\cite{ChakrabartiBook} The relationship between the FFTFIM on a regular lattice and
the QDM on the dual lattice has been first emphasized
by Moessner, Sondhi and Chandra\cite{moessner1} who showed (see also Ref. 1) that,
in the limit $\Gamma/J \rightarrow 0$, the FFTFIM on the honeycomb lattice
maps onto the QDM on the triangular lattice with $t = \Gamma^2/J$ and $V = 0$.
For the FFTFIM on the honeycomb lattice,
they also carried out a Landau-Ginzburg analysis and identified
four soft modes which, upon lowering $\Gamma/J$, simultaneously become
gapless, leading to a surprisingly large unit cell of 48 sites. Details
of this calculation have been reported later by Moessner and Sondhi in
Ref.~\onlinecite{moessner3}. These authors further conjectured that the
translational symmetry breaking transition out of the paramagnetic phase
coming from large $\Gamma/J$ provides a reasonable description of
the transition between the RVB phase and the intermediate phase of the QDM
on the triangular lattice.\cite{moessner2}

Building on this conjecture, Misguich and one of the present authors have
carried out a semiclassical investigation of the paramagnetic phase of the
FFTFIM on the honeycomb lattice\cite{misguich} and have shown that the
dispersion of the spin waves and their softening at the transition are in
remarkable agreement with the dispersion of visons in the QDM on the
triangular lattice and their crystallization transition as revealed by
quantum Monte Carlo (QMC) simulations.\cite{ralko3} However, the
analysis of Ref. \onlinecite{misguich} {has not covered the small
$\Gamma/J$ parameter range.}

In the present paper, we perform a systematic investigation of the FFTFIM
on the honeycomb lattice in the complete parameter range $0\le \Gamma/J <
+\infty$ with classical and semi-classical approximations. As we shall
see, the classical phase diagram is much richer than expected, with
an infinite number of
different crystalline phases below the paramagnetic phase: a plaquette
phase, a cascade of mixed phases, and a highly degenerate columnar phase.
Quantum fluctuations have been treated within linear spin-wave theory, leading to
a partial lifting of the degeneracy of the columnar phase, and to an
increase of the size of the region occupied by the plaquette phase.

To make contact between the physics of the FFTFIM and of the QDM, it is
useful to introduce a gauge theory defined on the triangular lattice by
the Hamiltonian:
\begin{equation}
H=- J \sum_l \tau^x_l  - \Gamma \sum_i \prod_{l(i)}
\tau^z_{l(i)}\,,\vspace*{-2mm}
\end{equation}
where $i$ runs over the sites of the dual honeycomb lattice, and $l(i)$
are the three bonds forming the triangular plaquette around site $i$. As
shown by Moessner, Sondhi and Fradkin,\cite{moessner4} the FFTFIM is
equivalent, up to a twofold degeneracy, to the odd sector of this gauge
theory defined by
\begin{equation}
\prod_{l[a]} \tau^x_{l[a]}=-1\,,\vspace*{-2mm}
\end{equation}
for all $a$, where $a$ is a site of the triangular lattice, and the
product over $l[a]$ runs over the six links emanating from $a$. For a
compact discussion of the correspondence between the three models, see
e.g. the introduction of Ref.~\onlinecite{misguich}.

\begin{figure}
 \centering
\includegraphics[width=5cm]{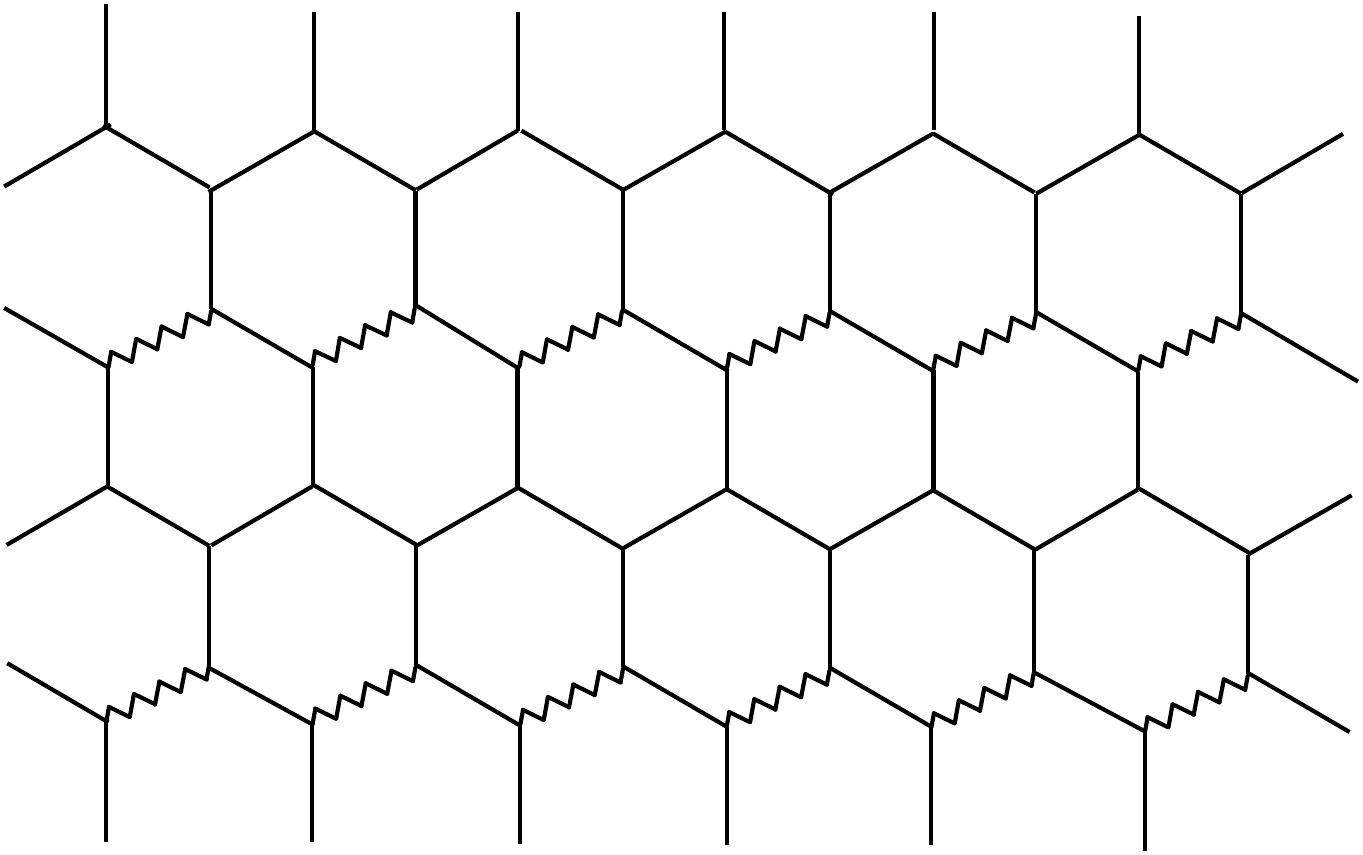}
\caption{Sketch of the gauge used in most of the paper. Here and
below antiferromagnetic bonds (with $M_{ij}=-1$) are shown by zigzags,
all other bonds being ferromagnetic (with $M_{ij}=+1$).} \label{fig:Choice
Of Gauge}
\end{figure}

The discussion of the ordered phases is simpler in the context of the
gauge theory. Indeed, in the FFTFIM language, the actual orientation of
the spins in a given state depends on the choice of the matrix $M_{ij}$.
By contrast, the dimer operator of the gauge theory defined by
\begin{equation}\label{eq:AverageDimerDensity}
d_{l}=\frac{1}{2} ( 1- \tau_l^x )\,, \vspace*{-2mm}
\end{equation}
translates into
\begin{equation}\label{eq:ClassicalAverageDimerDensity}
d_{ij}=\frac{1}{2}\left( 1-M_{ij}\frac{S_i^zS_j^z}{S^2} \right)\,,
\end{equation}
in the Ising language, and its expectation value does not depend on the
choice of $M_{ij}$. Another advantage of the gauge-invariant language is
that it allows to make a direct comparison with the numerical results
obtained on the QDM since they live on the same lattice and are defined in
terms of the same link operators. So, while all reasonings and
calculations will be performed in the context of the FFTFIM, the only
formulation adapted to the semiclassical approach, the structures of
different ordered phases will be also described in gauge-invariant terms.
Throughout the paper, we {use only gauges in which each hexagon of the
lattice contains exactly one antiferromagnetic bond (with $M_{ij}=-1$) and
five ferromagnetic bonds (with $M_{ij}=+1$).} Most results will be
presented for the simplest periodic arrangement of the antiferromagnetic
bonds shown in Fig.\ref{fig:Choice Of Gauge}. However, this choice of
gauge does not always lead to the smallest possible unit cell in terms of
the spin representation. Thus, we will also introduce other gauges
whenever this is helpful.

The paper is organized as follows. In section II, we concentrate on the
limit $\Gamma/J\ll 1$, which has not been considered in
Refs.\onlinecite{moessner1,moessner2,moessner3,misguich}, and we show that
columnar phases reminiscent of the $V\rightarrow -\infty$ limit of the QDM
are stabilized. In section III, we revisit the vicinity of the RVB phase.
We recover the symmetry predicted by the Landau-Ginzburg approach of
Refs.~\onlinecite{moessner1,moessner3} and by the spin-wave analysis of
Ref.~\onlinecite{misguich}, but we find that the bonds with the largest
dimer density form separate {4-site} rhombic plaquettes instead of having
a uniform distribution inside a 12-site unit cell as reported in
Ref.~\onlinecite{misguich}. The reasons for this discrepancy are explained
in subsection \ref{appendix:Difference Plaquette and Misguich Phases}. In
section IV, we discuss the transition between the plaquette phase and the
columnar phase and show that they are separated by
a region of intermediate phases of mixed character.
The stability of these phases with respect to
quantum fluctuations and the semi-classical phase diagram are discussed in
section V. The paper ends with a short conclusion in section VI.

%*****************************************************************************************************
%*****************************************************************************************************

\section{Columnar phase}

In this section we discuss the properties of the model when $\Gamma/J$ is
small. The argument proceeds in three steps. First we determine the ground
state manifold of the Heisenberg model with purely Ising-like interactions
in the absence of magnetic field ($\Gamma=0$). Then we investigate
how the extensive degeneracy of these ground states is lifted by a small
transverse field. Finally, we discuss the effect of quantum fluctuations
in the context of linear spin-wave theory.

\subsection{Zero transverse field}

In the absence of a transverse magnetic field ($\Gamma=0$), we are left
with a model without quantum fluctuations in which the interaction
term couples only the $z$ components of neighboring spins on the
honeycomb lattice. With our choice of gauge, one bond on each hexagon is
antiferromagnetic ($M_{ij}=-1$) and the others are ferromagnetic
($M_{ij}=1$). Frustration is present since it is clearly impossible to
minimize the energy of all bonds of a given hexagon.

For Ising spins, i.e. spins which can only point up or down along the $z$
direction, the best one can do is to satisfy five bonds leaving one
unsatisfied. This can be done in six different ways according to which
bond is not satisfied (``frustrated'') and the resulting energy is $-4J$.
Up to a global reversal of the spins, a ground state is characterized by
the distribution of frustrated bonds such that there is exactly one of
them per hexagon.

For three-dimensional vectors of norm $S$, the situation is slightly more
subtle because the twelve Ising configurations with all spins parallel or
antiparallel to $z$ axis are not the only ground states of a single
hexagon. To see this, let us consider a single hexagon and investigate the
possibility of a given spin $i$ not to be directed along $z$. The
variation of the energy of the hexagon
\begin{equation}                        \label{eq: single hexagon energy}
E_{\rm hex}=-J\sum_{j=1}^6{M_{j,j+1}\cos\theta_j\cos\theta_{j+1}}\,
\end{equation}
(where {the} angle $\theta_j$ parameterizes the deviation of spin $j$ from
the $z$ axis) with respect to $\theta_i$ leads to the condition
\begin{equation}               \label{eq: Hexagon energy minimization}
M_{i-1,i}\cos\theta_{i-1}+M_{i,i+1}\cos\theta_{i+1}=0\,.
\end{equation}
If this condition is satisfied, the terms in Eq. (\ref{eq: single hexagon
energy}) which depend on $\theta_i$ drop out, so that one is left with the
energy of an open chain of five spins. In an open chain one can trivially
minimize the energy of each bond by choosing
$\cos\theta_{j+1}=M_{j,j+1}\cos\theta_j=\pm 1$ which leads to $E=-4J$ and
{to the automatic fulfillment of condition (\ref{eq: Hexagon energy
minimization}), leaving $\theta_i$ arbitrary.} Note that this argument
excludes a deviation from {the} $z$ axis of more than one spin, since the
energy of a five-spin open chain cannot be as low as $-4J$ if not all five
spins are along $z$. So, for three-dimensional spins, the energy of a
single hexagon is minimal as soon as it is minimal for four consecutive
bonds, and the spin at the remaining site can have any direction.

It is natural to ask whether this additional freedom increases the
degeneracy of the ground state manifold of the {continuous model in
comparison with} the case of Ising spins. To demonstrate that this is not
the case, let us assume that at site $i$ the spin is not along $z$. To
minimize simultaneously the energy of the three hexagons to which it
belongs, three conditions of the form (\ref{eq: Hexagon energy
minimization}) must be fulfilled:
\begin{equation}   \label{eq:System_three_hexagons}
 \begin{array}{c}
  Y_1+Y_2=0\,,\\[1mm]
  Y_2+Y_3=0\,,\\[1mm]
  Y_3+Y_1=0\,,
 \end{array}
\end{equation}
where $Y_a=M_{i,i_a}\cos\theta_{i_a}=\pm 1$ (with $a=1,2,3$) and $i_a$ are
the three nearest neighbors of site $i$. It is evident that the
restriction $Y_a=\pm 1$ does not allow all three equations
(\ref{eq:System_three_hexagons}) to be satisfied simultaneously.
Therefore, it is impossible for any spin not to point along $z$, and the
ground state manifold coincides with that of the frustrated Ising model
with the same lattice, {i.e.} it consists of all Ising configurations with
one frustrated bond per hexagon. Each of these states is a local minimum
of the Hamiltonian.

\subsection{Classical ground states in small transverse field
\label{sec:classical columnar states}}

Let us now switch on a small transverse field and study how the
local minima of the classical Hamiltonian % found in the previous subsection
evolve {upon increasing the field}. Since the field is along $x$, the
spins are expected to acquire a small $x$ component, and to describe the
spin configuration evolving from a given ground state of the pure Ising
case, we use the parametrization:
\begin{equation}                                \label{eq:Classical Spins}
\begin{array}{l}
    S_i^x=S\sin\theta_i \,,\\[1mm]
    S_i^z=\sigma_iS\cos\theta_i \,,
\end{array}
\end{equation}
where $\sigma_i=\pm 1$ is the sign of $S_i^z$ and is determined by the
ground state of the pure Ising case around which we expand. In terms of
the gauge-invariant bond variable
$\tau_{ij}=M_{ij}\sigma_i\sigma_j$, which is equal to -1(+1) if the bond
$\langle i,j\rangle$ is frustrated (not frustrated),
%{\cmat $\Gamma=0$} ,
the classical energy can be rewritten as
\begin{equation}\label{eq:Classical energy}
 E=-{J}\sum_{\langle i,j\rangle}{ \tau_{ij}\cos\theta_i\cos\theta_j}-{\Gamma}\sum_{i}\sin\theta_i\,.
\end{equation}

In the limit $\Gamma\ll J$ the deviations from the $z$ direction are
small, and the classical energy can be expanded in the variables
$\theta_i$ around $\theta_i=0$. To second order, the interaction term
in Eq.~(\ref{eq:Classical energy}) decouples:
$\tau_{ij}\cos\theta_i\cos\theta_j \approx
\tau_{ij}(1-{\theta_i^2}/{2}-{\theta_j^2}/{2})$. Now, for any ground state
of the pure Ising case, the set $\{\tau_{ij}\}$ is such that only one bond
in each hexagon is frustrated. Therefore each site belongs
at most to one frustrated bond. % (then we call it
% frustrated site) or to no frustrated bonds (non frustrated site).
If we denote by F (resp. NF) the set of  what we call below frustrated
(resp. non frustrated) sites, namely, the sites belonging to one
frustrated bond (resp. no frustrated bond), the energy up to
second order can be rewritten:
\begin{equation}\label{eq:Classical energy second order split}
E^{(2)}=E_{\Gamma=0}+\sum_{i\in{\rm
F}}{\left(\frac{J}{2}{\theta_i^2}-\Gamma\theta_i\right)} +\sum_{i\in{\rm
NF}}{\left(\frac{3J}{2}{\theta_i^2}-\Gamma\theta_i\right)} \,.
\end{equation}
Minimizing $E^{(2)}$ with respect to $\{\theta_i\}$ leads to
\begin{equation}          \label{eq: theta_i}
\theta_i=\left\{\begin{array}{ll}
\Gamma/J & \mbox{ for } i\in\mbox{F}\,,
\\
\Gamma/3J & \mbox{ for } i\in\mbox{NF}\,.
\end{array}\right.
\end{equation}
Since the number of frustrated and non frustrated sites is the same for
all ground states, the energy up to second order in $\theta_i$
is the same in all ground states. So, second order corrections do not lift
the degeneracy. They only induce a difference in orientation between the
spins which belong to a frustrated bond and those which do not.

\begin{figure}[htbp]
\centering
 \subfigure[]{\label{fig:2sitesFrustrated} \includegraphics[height=2cm]{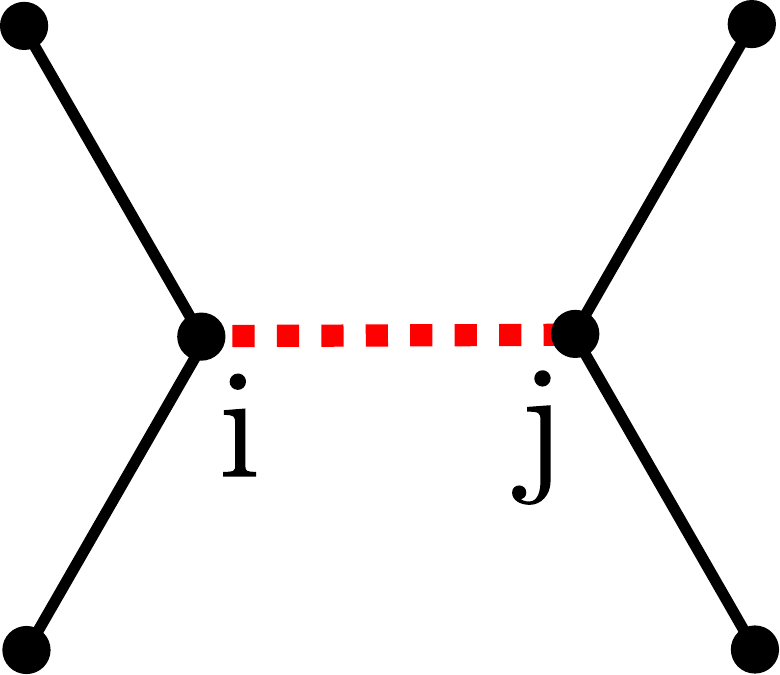}}
\hspace{1cm}
 \subfigure[]{\label{fig:2sitesNonFrustrated} \includegraphics[height=2cm]{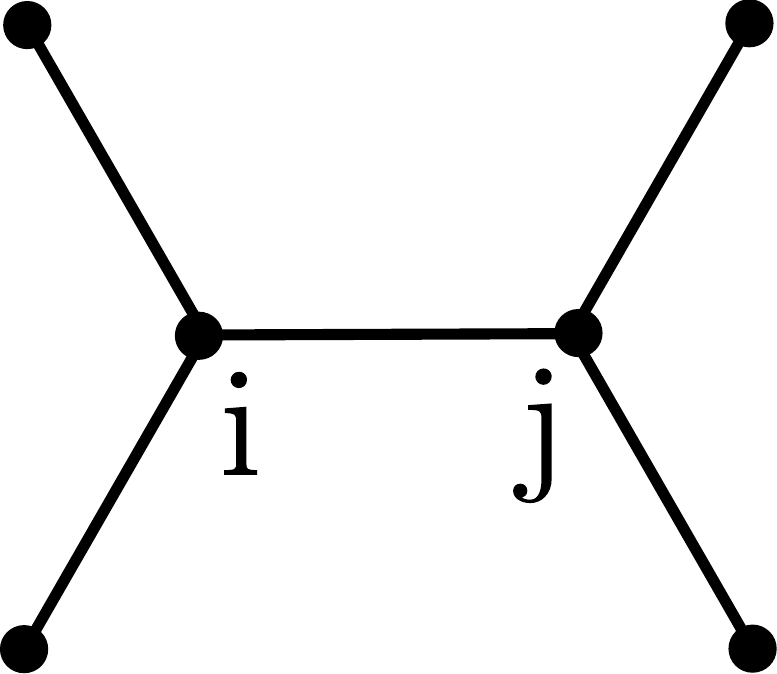}}\\
 \subfigure[]{\label{fig:2sitesSideFrustrated} \includegraphics[height=2cm]{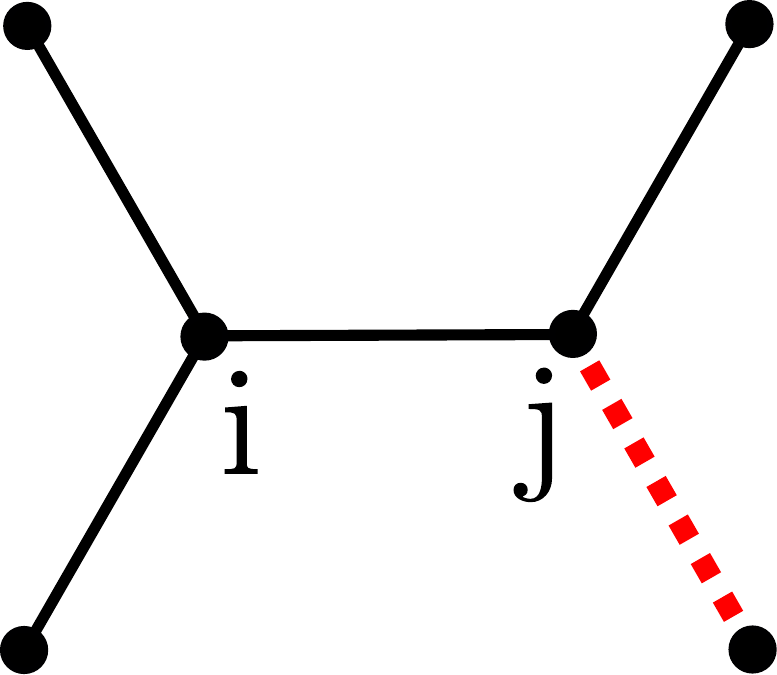}}
\hspace{1cm}
 \subfigure[]{\label{fig:2sitesDoublyFrustrated} \includegraphics[height=2cm]{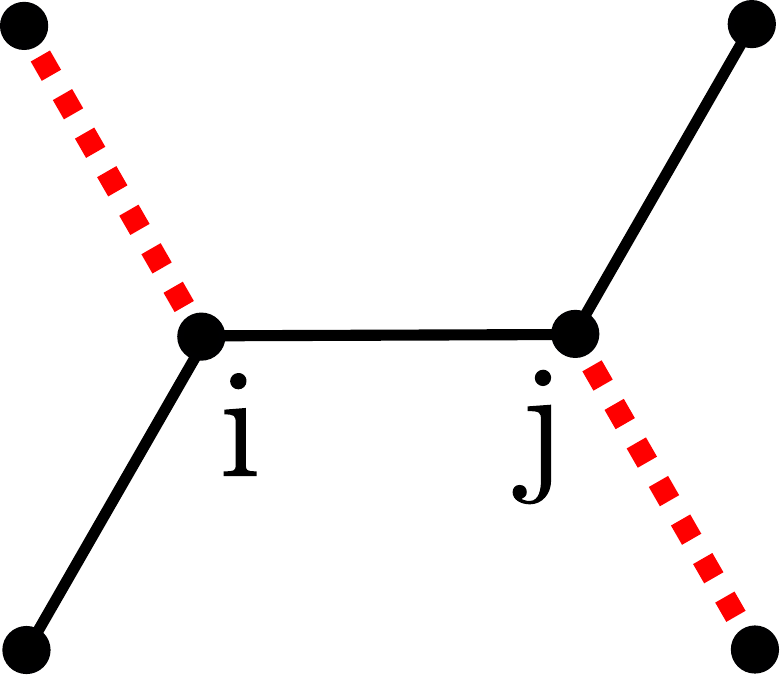}}
\caption{(Color online) Local configurations of frustrated bonds leading to different contributions of the fourth order crossed term $-\frac{J}{4}\tau_{ij}\theta_i^2\theta_j^2$.
% $N_a, N_b, N_c$ and $N_d$ denote the number of times each of the four configurations depicted above appear in a given classical state.
}\label{fig:Combining two sites}
\end{figure}

So to lift the degeneracy we have to push the expansion in $\theta_i$ to
higher orders. To fourth order, it reads:
\begin{equation}\label{eq:Energy to fourth order}
\begin{array}{ll}
E^{(4)}=&E_{\Gamma=0}
% \\ [2mm] &
+\displaystyle{\sum_{i\in{\rm F}}} {\left[
J\left(\frac{\theta_i^2}{2}-\frac{\theta_i^4}{4!}\right)-\Gamma
\left(\theta_i-\frac{\theta_i^3}{3!}\right)\right]}
\\ [4mm] &
+\displaystyle{\sum_{i\in{\rm
NF}}}{\left[3J\left(\frac{\theta_i^2}{2}-\frac{\theta_i^4}{4!}\right)-\Gamma
\left(\theta_i-\frac{\theta_i^3}{3!}\right)\right]}
\\ [4mm] &
-\frac{J}{4}\displaystyle{\sum_{\left\langle
i,j\right\rangle}}{\tau_{ij}\theta_i^2\theta_j^2}\,. \\ [-8mm]
\end{array}
\end{equation}
From the previous discussion, we know that the values of $\theta_i$
minimizing the energy to order $O(\theta^2)$ are given by Eq. (\ref{eq:
theta_i}).   Injecting these solutions into the fourth order expansion of
the energy, we notice that the terms $\theta_i^3$ and $\theta_i^4$ only
contribute in two different ways depending on the type of site (frustrated
or non frustrated). They will thus not lift the degeneracy. By contrast,
the crossed terms $\tau_{ij}\theta_i^2\theta_j^2$ contribute in four
different ways depending on the environment of the sites $i$ and $j$. The
four cases are illustrated in Fig.~\ref{fig:Combining two sites}.

The contributions of the fourth order crossed terms to the energy for the
different configurations in units of $\Gamma^4/4J^3$ are $+1$ for
\ref{fig:2sitesFrustrated}, $-\frac{1}{9}$ for
\ref{fig:2sitesSideFrustrated}, $-\frac{1}{81}$ for
\ref{fig:2sitesNonFrustrated} and $-1$ for
\ref{fig:2sitesDoublyFrustrated}. Since these energies are not equal,
these crossed terms are expected to lift the degeneracy, at least
partially.

For a lattice of $N_{\textrm{hex}}$ hexagons the total number of bonds is
$3 N_\textrm{hex}$. The constraint that each hexagon has one frustrated
bond implies that the number of frustrated bonds is equal to
$N_{\textrm{hex}}/2$. This fixes the number $N_a$ of configurations
\ref{fig:2sitesFrustrated} to be equal to $N_\textrm{hex}/2$. By contrast,
the numbers of configurations of type \ref{fig:2sitesNonFrustrated},
\ref{fig:2sitesSideFrustrated} and \ref{fig:2sitesDoublyFrustrated}
(respectively $N_b,N_c,N_d$) depend on the way the frustrated bonds are
arranged on the lattice. However $N_b,N_c$ and $N_d$ are not independent
but have to satisfy the following relations:
\begin{equation}\label{eq:First Relation NB NC ND}
 N_b+N_c+N_d=\frac{5}{2}N_\textrm{hex}\,,
\end{equation}
\begin{equation}\label{eq:Second Relation NB NC ND}
N_a=\frac{1}{4}(N_c+2N_d)\,.
\end{equation}
Eq.~(\ref{eq:First Relation NB NC ND}) comes from the conservation of the
total number of bonds $N_a+N_b+N_c+N_d=3 N_\textrm{hex}$, whereas the
right-hand side of Eq.~(\ref{eq:Second Relation NB NC ND}) comes from
counting all frustrated bonds by looking at how many of them are adjacent
to each of the non frustrated bonds. The result of this calculation has to
be divided by four, because in such a procedure each frustrated bond is
counted four times.

\begin{figure}[htbp]
    \centering
\subfigure[ $~1^\textrm{st}$ columnar state]{\label{fig:First Columnar
State} \includegraphics[width=3.8cm]{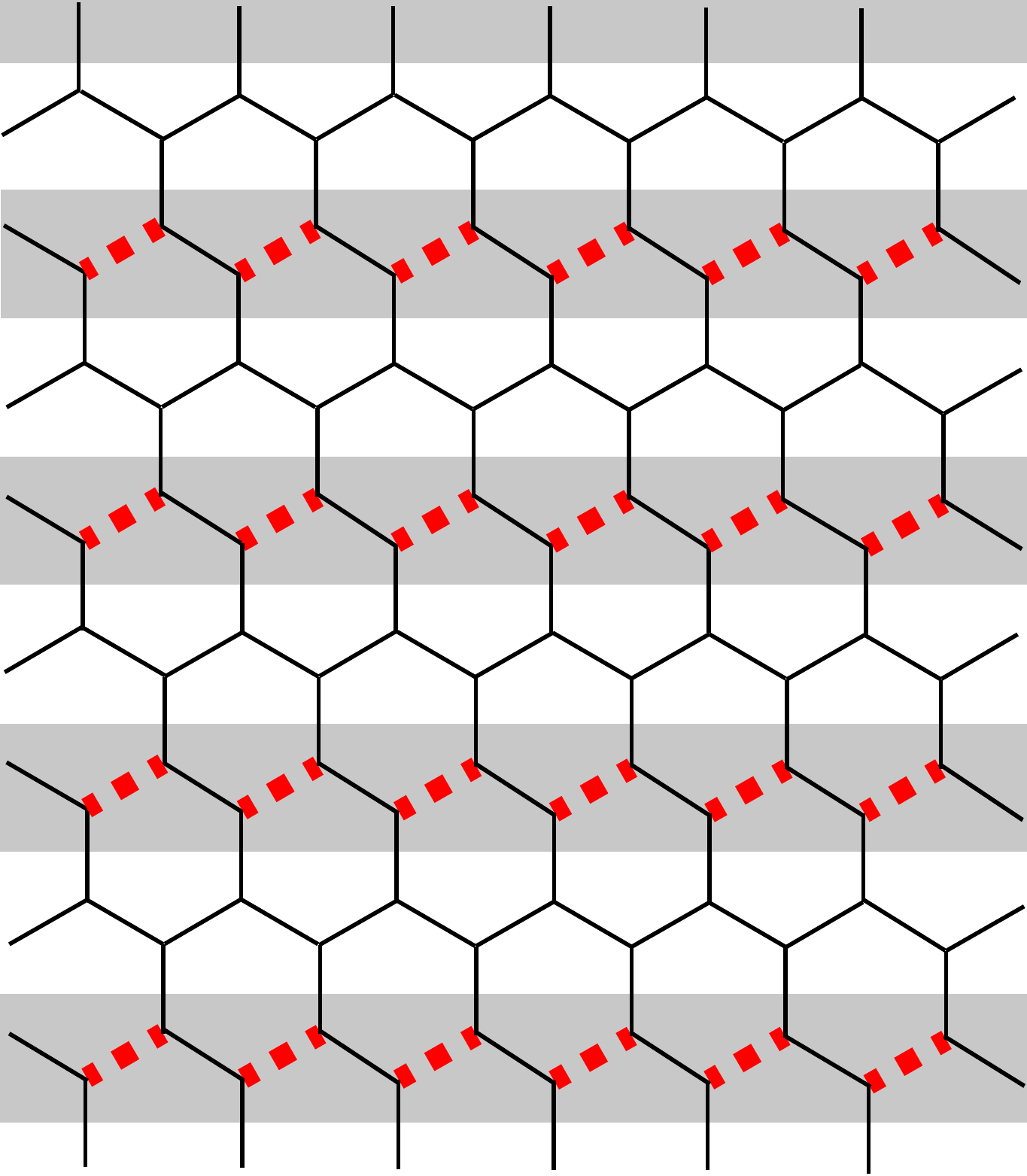}} \hspace{0.5cm}
\subfigure[ $~2^\textrm{nd}$ columnar state ]{\label{fig:Second Columnar
State} \includegraphics[width=3.8cm]{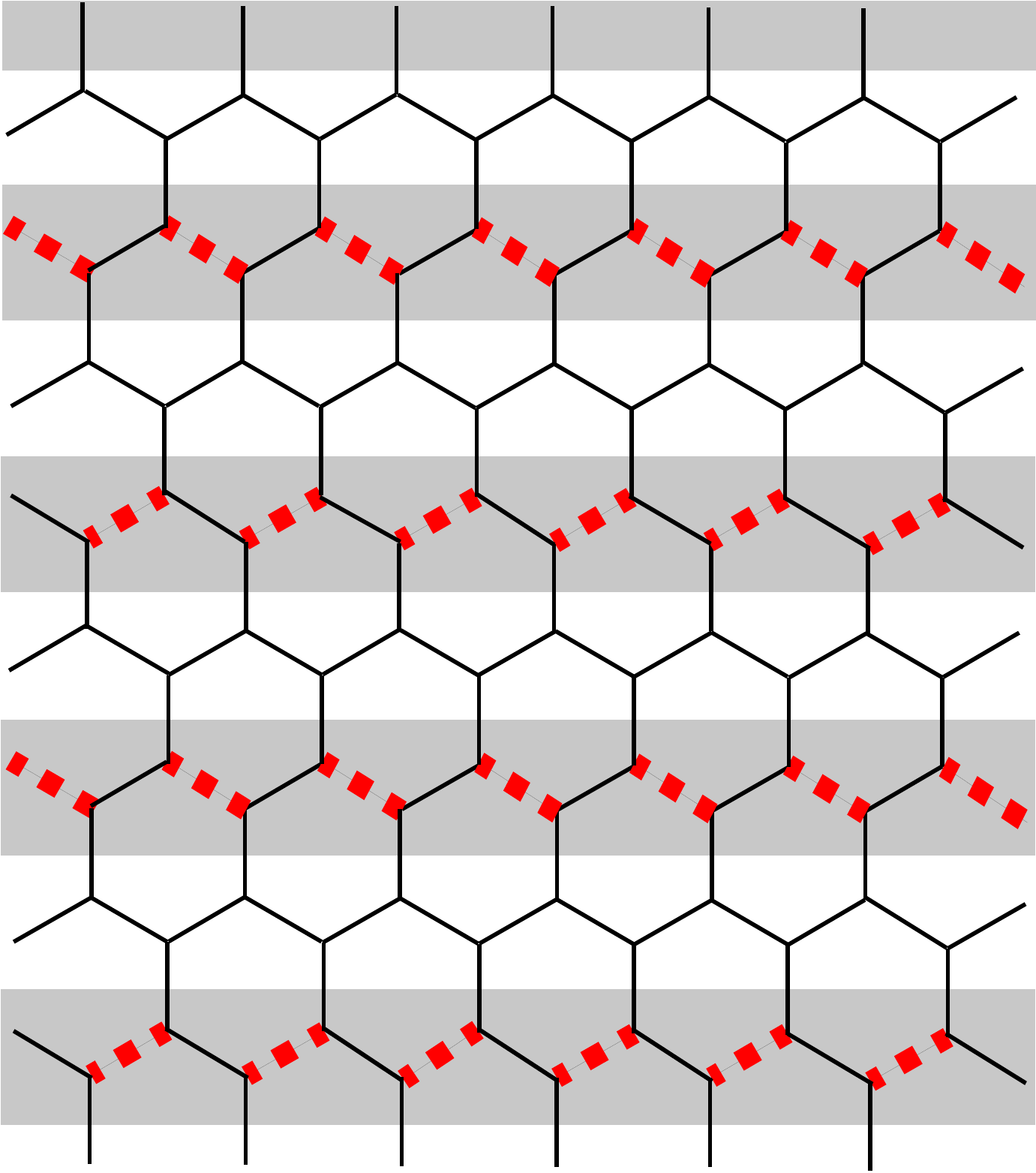}} \subfigure[
$~3^\textrm{rd}$ columnar state]{\label{fig:Third Columnar State}
\includegraphics[width=3.8cm]{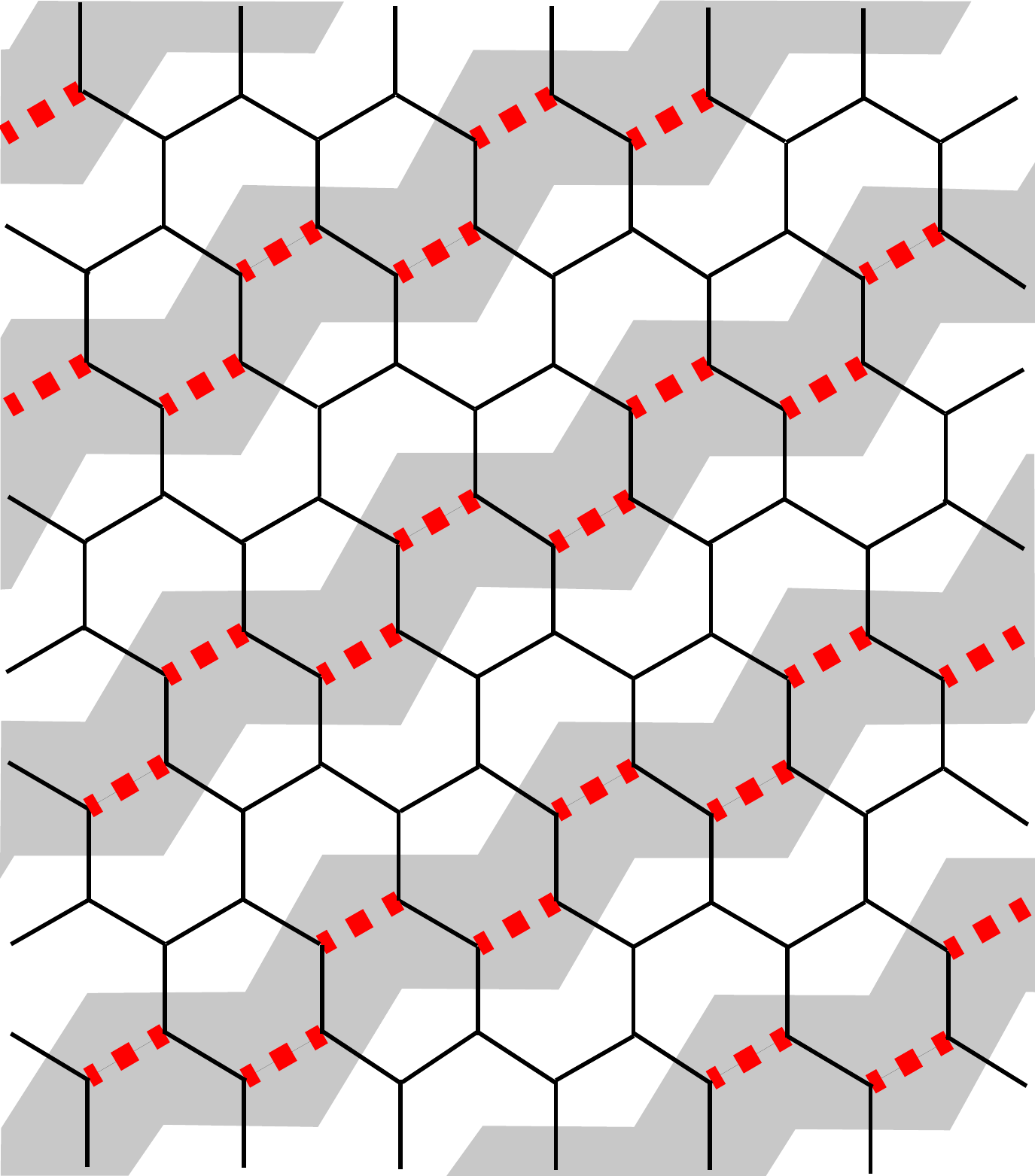}} \hspace{0.5cm}
\subfigure[ {$~4^\textrm{th}$ columnar state}]{\label{fig:3rdColumnar
state half domain walls}
\includegraphics[width=3.8cm]{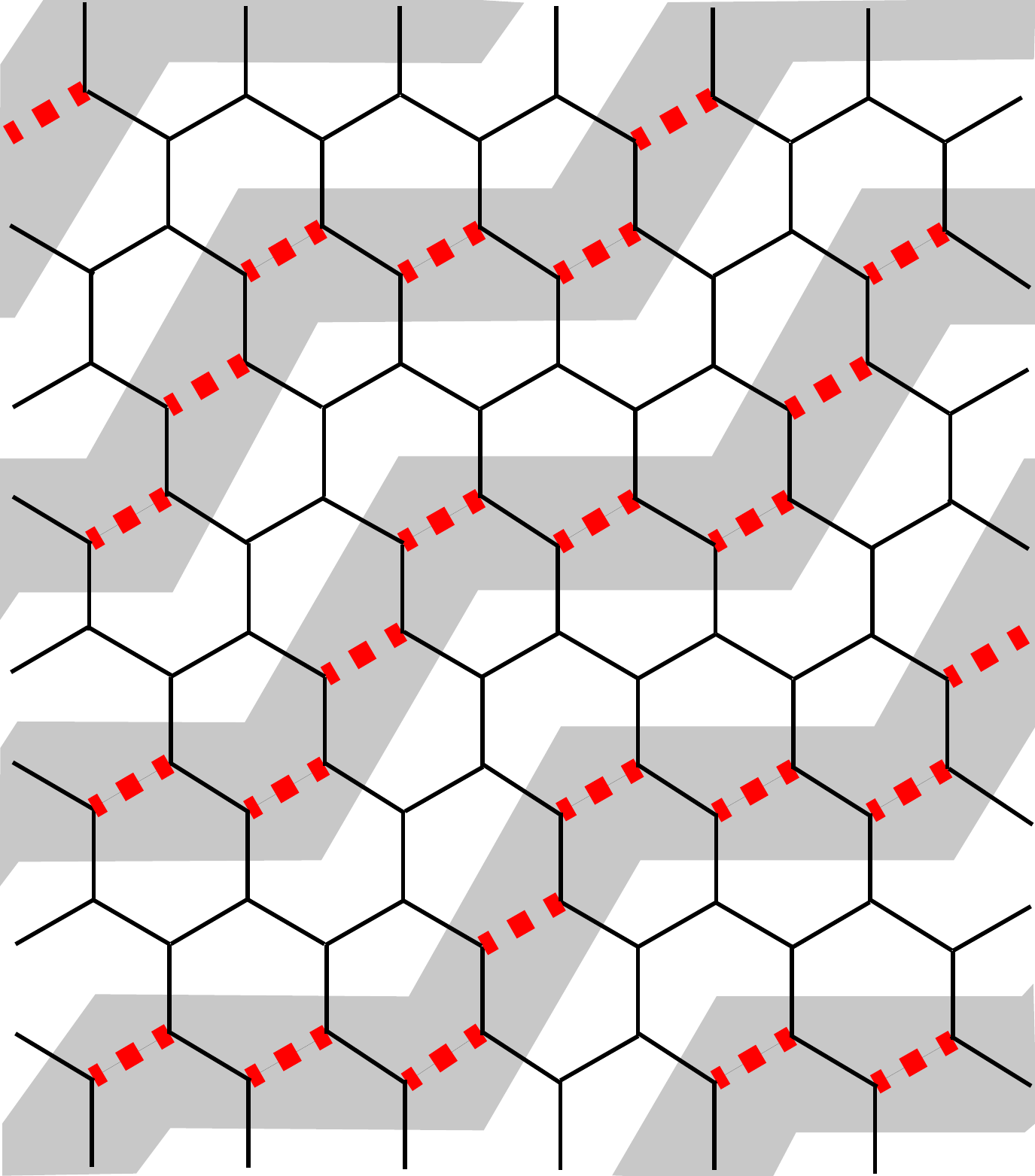}} \caption{(Color
online) Examples of columnar states. The frustrated bonds are represented
as dashed red lines. In the dimer representation, the bonds of the dual
triangular lattice which intersect the frustrated bonds of the
honeycomb lattice have the highest dimer density. The $4^\textrm{th}$
columnar state differs from the $3^{\textrm{rd}}$ one by having exactly
half the  number of domain walls of the second type (see main text).}
\label{fig:DifferentColumnarStates}
\end{figure}

The total contribution of the fourth order crossed terms of the energy can
then be written as:
\begin{equation}\label{eq: Fourth order contrib of the energy Columnar States}
\begin{array}{rcl}
 -\frac{J}{4}\sum_{\left\langle i,j\right\rangle}{\tau_{ij}\theta_i^2\theta_j^2} &\approx&
          \frac{J}{4}\left(\frac{\Gamma}{J}\right)^4\left(N_a - N_d -\frac{1}{9}N_c -\frac{1}{81}N_b\right) \\ [3mm]
 &\approx&\frac{J}{4}\left(\frac{\Gamma}{J}\right)^4\left(\frac{22}{81}{N_{\textrm{hex}}} - \frac{64}{81}N_d  \right)\,.
\end{array}
\end{equation}
This contribution is a decreasing function of $N_d$, so the lowest
energy will be reached for the largest possible value of $N_d$. Now,
since there is only one frustrated bond per hexagon, $N_d$ cannot exceed
the number of frustrated bonds, $N_a={N_\textrm{hex}}/{2}$. This
upper limit is reached for configurations in which all the frustrated
bonds are organized into chains of alternating frustrated and non
frustrated bonds (see examples in Fig.~\ref{fig:DifferentColumnarStates}).
In what follows we refer to this family of states as columnar states (see
Fig.~\ref{fig:DifferentColumnarStates}). In columnar states Eqs.
(\ref{eq:First Relation NB NC ND}) and (\ref{eq:Second Relation NB NC ND})
fix both $N_b$ and $N_c$ to be equal to $N_{\textrm{hex}}$.

So, the fourth order contribution to the energy partially lifts the
degeneracy and selects the family of columnar states. A priori, higher
orders might further lift the degeneracy. That this is not the case is
best seen by constructing the exact local minima that correspond to
columnar states. We start by rewriting the energy:
\begin{equation}\label{eq:Full energy split}
\begin{array}{rl}
 E=
& \displaystyle{-\sum_{i\in{\rm NF}}\left[{\frac{J}{2}\cos\theta_i
\left(\cos\theta_{i_1}+\cos\theta_{i_2}+\cos\theta_{i_3}\right)+\Gamma\sin\theta_i}\right]}
\\ [3mm]
-&\displaystyle{\sum_{j\in{\rm F}}\left[{\frac{J}{2}\cos\theta_j
\left(-\cos\theta_{j_1}+\cos\theta_{j_2}+\cos\theta_{j_3}\right)+\Gamma\sin\theta_j}\right]}\,,
\end{array}
\end{equation}
where $i_1,i_2,i_3$ (resp. $j_1,j_2,j_3$) are the three neighbors of site
$i$ (resp. $j$), and the frustrated bond is taken to be between sites $j$
and $j_1$. To minimize the energy, the set of angles
$\{\theta_i,\theta_j\}$ must be a solution of the equations:
\begin{equation}\label{eq:General partial derivatives}
\begin{array}{l}
 \frac{\partial E}{\partial \theta_i}=J\sin\theta_i\left(  \cos\theta_{i_1}
 +\cos\theta_{i_2}+\cos\theta_{i_3} \right) -\Gamma\cos\theta_i = 0\,,
 \\ [3mm]
 \frac{\partial E}{\partial \theta_j}=J\sin\theta_j\left( -\cos\theta_{j_1}
 +\cos\theta_{j_2}+\cos\theta_{j_3} \right) -\Gamma\cos\theta_j = 0\,.
\end{array}
\end{equation}
Now, in columnar structures, all frustrated sites have identical
environments (with exactly two frustrated neighbors) and all unfrustrated
sites also have identical environments (with exactly one frustrated
neighbor). So, if the angles $\theta_1$ and $\theta_2$ satisfy the
equations:
\begin{equation}\label{eq:Partial derivatives columnar states}
\begin{array}{l}
 J\sin\theta_1\left(2\cos\theta_1+\cos\theta_2 \right) -\Gamma\cos\theta_1 = 0\,,\\ [3mm]
J\sin\theta_2 \cos\theta_1 -\Gamma\cos\theta_2 = 0\,,
 \end{array}
\end{equation}
then the set of angles
\begin{equation}
\label{eq:theta=theta1 or theta2}
    \theta_i=\left\{\begin{array}{ll}
      \theta_1 & \mbox{for } i\in\mbox{NF} \\
      \theta_2 & \mbox{for } i\in\mbox{F}
    \end{array}\right.
\end{equation}
% $\theta_i=\theta_1$ if $i$ is non frustrated and $\theta_i=\theta_2$ if
% $i$ is frustrated
is a solution of Eqs.~(\ref{eq:General partial derivatives}). The non
trivial solutions of Eqs.~(\ref{eq:Partial derivatives columnar states})
describing the evolution of columnar states with the change of $\Gamma/J$
are given by:
\begin{equation}\label{eq:Classical Angles}
\begin{array}{l}
    \sin\theta_1=\frac{\sin\left(\beta/3\right)}{\cos\left(\beta\right)}\;,
\hspace{1cm}
    \sin\theta_2=\frac{\sin\left(\beta\right)}{\cos\left(\beta/3\right)}\;,
\end{array}
\end{equation}
where $\tan\beta=\Gamma/J$.

{The substitution} of Eq.~(\ref{eq:theta=theta1 or theta2}) into
Eq.~(\ref{eq:Classical energy}) shows that the classical energy of a
columnar state is given by
\begin{equation}\label{eq:Energy Columnar States}
 E_{\rm col}=-\frac{N_\textrm{}}{2}\left[J\cos{\theta_1}\left(\cos{\theta_1}+\cos{\theta_2}\right)
 +\Gamma(\sin{\theta_1}+\sin{\theta_2}) \right]\,,
\end{equation}
where $N$ is the total number of sites. Naturally, the variation of
$E_{\rm col}$ % Eq. (\ref{eq:Energy Columnar States})
with respect to $\theta_1$ and $\theta_2$ reproduces Eqs. (\ref{eq:Partial
derivatives columnar states}) which we used to find the values of
$\theta_1$ and $\theta_2$. In order to verify
% the {stabilization} of columnar states over other minima of classical
% energy persists even when the ratio $\Gamma/J$ is not small,
that it never becomes more advantageous to minimize $N_d$ rather then to
maximize it, we have studied also the solutions with $N_d=0$ and checked
that for any relation between $\Gamma$ and $J$ they have higher energy
than the columnar states (see Appendix \ref{appendix:staggered}).

A convenient classification of columnar states can be introduced by
describing them in terms of zero-energy domain walls formed on the
background of the simplest columnar state, an example of which is shown in
Fig.~\ref{fig:First Columnar State}. Below we call it the $1^\textrm{st}$
columnar state. In this state all frustrated bonds have the same
orientation and form straight columns shown in the figure by the shading.
% In the construction of columnar states we identify two types of
% domain walls.
In terms of Fig. \ref{fig:DifferentColumnarStates} the walls of the first
type are horizontal and take place whenever the orientation of the
frustrated bonds changes from left to right. The $2^{\textrm{nd}}$
columnar state (Fig.~\ref{fig:Second Columnar State}) corresponds to the
configuration having the highest possible density of such domain
walls.

The domain walls of the second type are perpendicular to the
frustrated bonds, and correspond to changing the orientation not of
frustrated bonds but of columns. The $3^{\textrm{rd}}$ columnar state
(Fig.~\ref{fig:Third Columnar State}) is the configuration having the
highest possible density of walls of the second type as the
orientation of the columns changes at every frustrated bond. Other
columnar states having the same classical energy can be obtained by
introducing arbitrary sequences of parallel domain walls either of the
first or of the second type separating domains of the $1^{\textrm{st}}$
columnar state. An analogous classification of columnar states had earlier
been introduced by Moessner and Sondhi\cite{moessner3} in terms of the
QDM.

\begin{figure}[htbp]
\centering
\subfigure[ $\textrm{ }1^{\textrm{st}}$ Columnar state: dimer representation]
{\label{fig:Columnar12x4} \includegraphics[height=2.8cm]{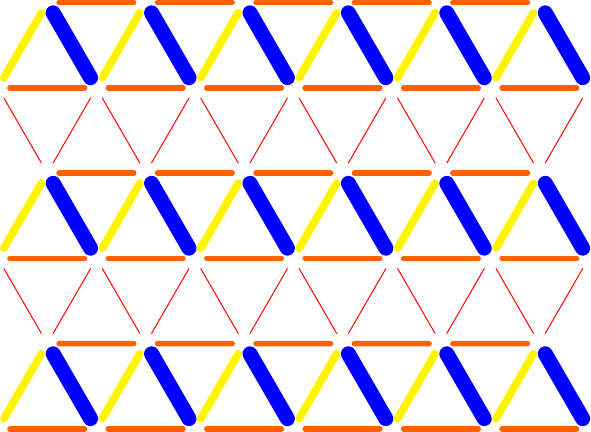}}\\
\vspace{5mm} \subfigure[ $\textrm{ }1^{\textrm{st}}$ Columnar state: spin
representation]{\label{fig:ColumnarSz}
\includegraphics[height=4.2cm]{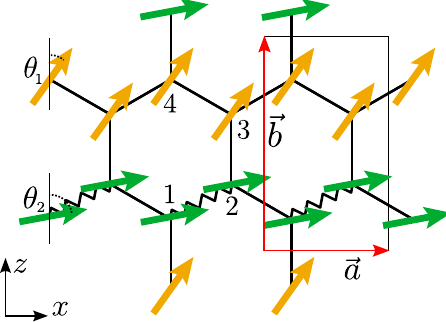}} \caption{(Color online)
(a) Plot of the dimer density $d_{ij}$ at $\Gamma/J=1.5$ for the $1^{\textrm{st}}$
columnar state. The thickness of the bonds is
proportional to $d_{ij}$. The dark blue bonds corresponding to the highest dimer
density are organized into columns. \\
(b) Spin configuration % on the honeycomb lattice
in the $1^{\textrm{st}}$ columnar state in the gauge of
Fig.~\ref{fig:Choice Of Gauge} (with the same notation for
antiferromagnetic bonds). The two types of arrows correspond to the two
{spin orientations} realized in that state. The unit cell is defined by
the two vectors $\vec{a}$ and $\vec{b}$ with
$|\vec{b}|=\sqrt{3}|\vec{a}|$.} \label{fig:ColumnarPhase}
\end{figure}

%\subparagraph{}
Fig.~\ref{fig:Columnar12x4} presents a plot of the dimer density for
the $1^{\textrm{st}}$ columnar state at $\Gamma/J=1.5$. The bonds of the
dual triangular lattice having the highest dimer densities are organized
into a columnar pattern. Fig.~\ref{fig:ColumnarSz} is a plot of the
$1^{\textrm{st}}$ columnar state in the classical spin model. The $S^z$
component of the spin on frustrated sites (green arrows in
Fig.~\ref{fig:ColumnarSz}) is smaller than that on non frustrated sites.

\subsection{Quantum fluctuations
\label{sec: Quantum fluctuations in columnar states}}

The effect of quantum fluctuations on the columnar states, in particular
their local stability and their degeneracy, has been investigated in the
context of linear spin-wave theory (LSWT). It is impossible to perform a
LSWT calculation for all columnar states since the family is infinite and
contains many members which are not periodic. The logic we have followed
is based on the expectation that the difference in energy between each
pair of states is determined primarily by the difference in the number of
domain walls they contain.

In Sec. \ref{sec:classical columnar states} we have established that
the structure of columnar solutions is described by
Eqs.~(\ref{eq:Classical Spins}),
%\begin{equation}\label{eq:Columnar solution}
% \begin{array}{ccl}
% S_i^x&=&S\sin\theta_i\,,\\[1mm]
%  S_i^z&=&\sigma_i S\cos\theta_i\,,
% \end{array}
% \end{equation}
where \makebox{$\sigma_i=\pm 1$} is determined by the ground state of the
pure Ising case and the values of the variables $\theta_i$ are given by
Eqs.~(\ref{eq:theta=theta1 or theta2}) and (\ref{eq:Classical Angles}). It
is convenient to start the construction of the Hamiltonian describing the
harmonic fluctuations around these states by performing a rotation of the
spins on each site,
\begin{equation}\label{eq:Local Spin Rotation}
\begin{array}{ccl}
 S_i^x&=&\sigma_i\cos\theta_i S_i^{x\prime}+\sin\theta_i S_i^{z\prime}\,,\\[1mm]
 S_i^y&=&S_i^{y\prime}\,,\\[1mm]
 S_i^z&=&-\sin\theta_i S_i^{x\prime}+\sigma_i\cos\theta_i S_i^{z\prime}\,.
\end{array}
\end{equation}
in such a way that the Hamiltonian expressed in terms of the
variables $S^{x\prime}$ and $S^{z\prime}$ has a ferromagnetic ground
state.

Mapping the new spin operators to Holstein-Primakoff bosons in the
harmonic limit: \cite{HolsteinPrimakoffTransformation}
\begin{equation}\label{eq:HP transformation}
 S_i^{z^\prime}=S-a_i^\dagger a_i\,, \hspace{1cm} S_i^{x^\prime}\approx\sqrt{\frac{S}{2}}\left(a_i+a_i^\dagger\right)\,,
\end{equation}
then yields the quadratic Hamiltonian:
\begin{eqnarray}\label{eq:H after HP transformation}
 H & = & E_{\rm col}+\displaystyle\gamma_1\sum_{i\in {\rm NF}}
 %\displaystyle\left[\frac{J}{S}\cos\theta_1\cos\theta_2+\frac{\Gamma}{S}\sin\theta_2\right]
 a_i^\dagger a_i %\\ [4mm] &
 +\displaystyle\gamma_2\sum_{i\in{\rm F}}
 %\displaystyle\left[\frac{J}{S}\cos\theta_1(2\cos\theta_1+\cos\theta_2)+\frac{\Gamma}{S}\sin\theta_1\right]
 a_i^\dagger a_i \\  & &
 -\frac{J}{2S}\displaystyle\sum_{\langle i,j\rangle}{M_{ij}\sin\theta_i
 \sin\theta_j \left[a_ia_j+a_i^\dagger a_j+\mbox{h.c.}\right]}\,,\nonumber
\end{eqnarray}
where
\begin{eqnarray}                                       \label{gamma1and2}
\gamma_1 & = &
({1}/{S})[J\cos\theta_1(2\cos\theta_1+\cos\theta_2)+\Gamma\sin\theta_1]\,,~~\\
\gamma_2 & = & ({1}/{S})(J\cos\theta_1\cos\theta_2+\Gamma\sin\theta_2)\,,~
\end{eqnarray}
and $E_{\textrm{col}}^{\textrm{}}$ is the classical energy of a columnar
state. Eq.~(\ref{eq:H after HP transformation}) can be reduced to a
gauge-invariant form (with $M_{ij}$ replaced by $\tau_{ij}$) by replacing
$a_i$ by $\sigma_i a_i$ and $a_i^\dagger$ by $\sigma_i a_i^\dagger$ in
Eqs.~(\ref{eq:HP transformation}). However, we use Eq.~(\ref{eq:H after HP
transformation})
% with the gauge of Fig. \ref{fig:Choice Of Gauge},
in the following because it allows an easy proof that domain walls of the
first type do not change the energy of the harmonic fluctuations.

It is evident that for $\theta_i$ given by Eq.~(\ref{eq:theta=theta1 or
theta2}), the expression in the right-hand side of Eq.~(\ref{eq:H after HP
transformation}) is exactly the same for all columnar states having the
same sets of frustrated and non frustrated sites. Since the introduction
of domain walls of the first type interchanges only the positions of
frustrated and non frustrated bonds forming straight columns, but does not
change the positions of frustrated sites [see Fig. \ref{fig:First Columnar
State} and Fig. \ref{fig:Second Columnar State}], the expression in the
right-hand side of Eq.~(\ref{eq:H after HP transformation}) will be
exactly the same for all columnar states which can be transformed one into
another by the introduction of some number of domain walls of the first
type. This proves that the contribution of the harmonic fluctuations to
the energy is the same for all members of the family of columnar states
having only domain walls of the first type.

After partitioning the honeycomb lattice into four sublattices in
accordance with the structure of the unit cell shown in Fig.
\ref{fig:ColumnarSz} and performing on each sublattice the  Fourier
transformation with wavevector $\vec q$, the quadratic bosonic Hamiltonian
% (\ref{eq:H after HP transformation})
of the $1^{\textrm{st}}$ columnar state is reduced to the form
\begin{equation}                           \label{eq:harmonic hamiltonian}
 H=E_{\textrm{col}}^{\textrm{}}
 + N \sum_{\vec{q}}{[\vec{a}^{\dagger}_{\vec{q}}\hat{H}(\vec{q})\vec{a}_{\vec{q}}-(\gamma_1+\gamma_2)]}
\end{equation}
In this expression, $\vec{a}^{}_{\vec{q}}$ is an eight-component vector
%\vec{a}^{\dagger}_{\vec{q}}\equiv
\makebox{$( a_{-\vec{q},1},a_{-\vec{q},2}, a_{-\vec{q},3},a_{-\vec{q},4},
a_{\vec{q},1}^{\dag},a_{\vec{q},2}^{\dag},a_{\vec{q},3}^{\dag},a_{\vec{q},4}^{\dag}
)$,} where $a_{\vec{q},n}$ are the bosonic operators with wavevector
$\vec{q}$ acting on the $n^{\textrm{th}}$ sublattice,
% site of the unit cell (figure~\ref{fig:ColumnarSz}),
%are the constants in diagonal terms in Eq. (\ref{eq:H after HP transformation}),
and $\hat{H}(\vec{q})$ is an $8\times8$ hermitian matrix
given by:
\begin{equation}                   \label{eq: 1st Columnar state matrix Mk}
% \begin{array}{l}
  \hat{H}(\vec{q})=\frac{1}{2}
             \left(\begin{array}{cccccccc}
             \gamma_2         &    \mu_{}    &        0         &     \delta_{}    &        0          &     \mu_{}    &        0         & \delta_{}  \\
             \mu_{}^{\star}   &   \gamma_2   &      \tau        &         0        &   \mu_{}^{\star}  &        0      &      \tau        &  0        \\
                     0        &   \tau_{}    &     \gamma_1     & \eta_{}^{\star}  &        0          &     \tau_{}   &        0         & \eta_{}^{\star}  \\
             \delta_{}^{\star}&         0    &      \eta_{}     &     \gamma_1     & \delta_{}^{\star} &       0       &     \eta_{}      &     0     \\
                     0        &    \mu_{}    &        0         &    \delta_{}     &     \gamma_2      &     \mu_{}    &        0         & \delta_{}  \\
             \mu_{}^{\star}   &         0    &     \tau_{}      &         0        &   \mu_{}^{\star}  &   \gamma_2    &     \tau_{}      &  0        \\
                     0        &   \tau_{}    &        0         & \eta_{}^{\star}  &        0          &     \tau_{}   &     \gamma_1     & \eta_{}^{\star}  \\
             \delta_{}^{\star}&         0    &     \eta_{}      &         0        & \delta_{}^{\star} &       0       &      \eta_{}     & \gamma_1 \\
        \end{array}\right)
\end{equation}
where
\begin{equation}
   \begin{array}{l}
%    \gamma_1=\frac{JG}{S}+\frac{\Gamma}{S}\sin\theta_2\,, \\[3mm]
%    \gamma_2=\frac{2JI}{S}+\frac{JG}{S}+\frac{\Gamma}{S}\sin\theta_1\,, \\[3mm]
 \mu\equiv\mu({\vec{q}})=-\frac{J\sin^2\theta_2}{2S}(-1+e^{i\vec{q}\vec{a}})\,, \\[3mm]
 \eta\equiv\eta({\vec{q}})=-\frac{J\sin^2\theta_1}{2S}(1+e^{i\vec{q}\vec{a}})\,, \\[3mm]
\delta\equiv\delta({\vec{q}})=\tau e^{i\vec{q}\vec{b}}\,,
~~~ %\\[3mm]
\tau= -\frac{J\sin\theta_1\sin\theta_2}{2S}\,.
\end{array}
%   \begin{array}{r}
%    G=\cos\theta_1\cos\theta_2\,,\\ [3mm]
%    I=\cos^2\theta_1\,,\\ [3mm]
%    A=\sin^2\theta_2\,,\\ [3mm]
%    H=\sin^2\theta_1\,,\\ [3mm]
%    D=\sin\theta_2\sin\theta_1\,,\\ [3mm]
%   \end{array}
%\end{array}
\end{equation}
{The vectors $\vec a$ and $\vec b$ are} shown in Fig.
\ref{fig:ColumnarSz}.

As discussed above, the harmonic Hamiltonian is the same for the whole
family of columnar states constructed by introducing an arbitrary number
of domain walls of the $1^{\textrm{st}}$ type. This family includes for
instance the $2^{\textrm{nd}}$ columnar state. In the harmonic
approximation, all these states have the same quantum corrections to the
energy, therefore to order $1/S$ the degeneracy is not lifted. Note
however that the absence of degeneracy lifting for this family of states
at the harmonic level is not related to a symmetry of the original
Hamiltonian. So we expect this degeneracy to be removed if one goes beyond
the harmonic approximation, and higher order terms are expected to select
either the $1^{\textrm{st}}$ or the $2^{\textrm{nd}}$ columnar state
depending on whether the energy of a domain wall of the first type is
positive or negative. However the effect of anharmonicities has not been
investigated in this work. Note that a similar effect, namely the
incapacity of harmonic fluctuations to fully lift a well-developed
accidental degeneracy of the ground states, has already been reported for
various other models (in particular, with {\em kagom\'{e}},
\cite{harris,chalker,ritchey}
honeycomb,\cite{KorshunovDoucotPhysRevLett.93.097003}
dice\cite{Korshunov05} and pyrochlore\cite{henley} lattices).

By contrast, the $3^{\textrm{rd}}$ columnar state is described by a
different harmonic Hamiltonian which is not written down here explicitly
because the number of sites per unit cell, hence the linear dimension of
the matrix $\hat{H}(\vec{q})$, is twice as large, so that the matrix
$\hat{H}(\vec{q})$ is $16\times 16$. The energy of zero point fluctuations
in this state turns out to be higher than in the $1^{\textrm{st}}$
columnar state (see Fig.~\ref{fig:Comparison of quantum fluctuations}).
This suggests that domain walls of the second type have a positive energy.

To support this statement, we have applied the same reasoning as used in
Ref. \onlinecite{KorshunovVallatBeckPhysRevB.51.3071} for the
investigation of the frustrated $XY$ model on a triangular lattice and
have considered the $4^\textrm{th}$ columnar state
(Fig.~\ref{fig:3rdColumnar state half domain walls}) which differs from
the $3^{\textrm{rd}}$ one {in that the density of domain walls of the
second type is exactly half as large}. Fig.~\ref{fig:Comparison of quantum
fluctuations} compares the numerically calculated differences between the
value of the quantum corrections to the energies of the $2^\textrm{nd}$,
$3^\textrm{rd}$ and $4^\textrm{th}$ columnar states and its value for the
$1^\textrm{st}$ columnar state. In particular, the inset in
Fig.~\ref{fig:Comparison of quantum fluctuations} presents the ratio of
these quantities for the $3^\textrm{rd}$ and $4^\textrm{th}$ states. This
ratio is very close to two, supporting the suggestion that the fluctuation
induced corrections to the energy are essentially proportional to the
density of domain walls of the second type.

\begin{figure}[htbp]
\includegraphics[width=8cm]{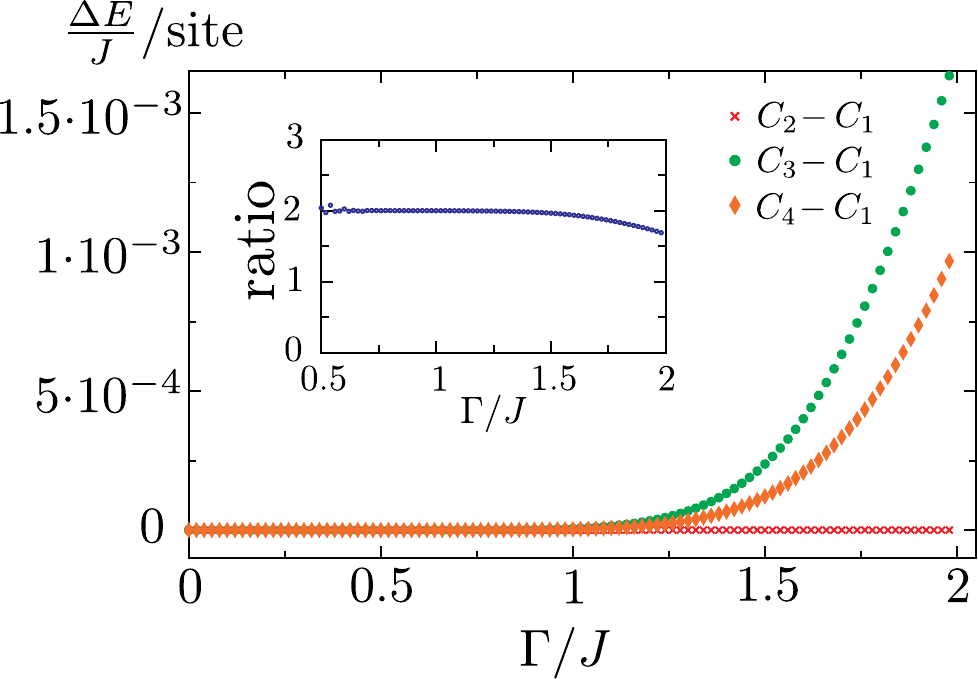}
\caption{(Color online) Energies (per site) of the $2^\textrm{nd}$
(red crosses), $3^\textrm{rd}$ (green circles) and $4^\textrm{th}$ (orange
diamonds) columnar states calculated in the harmonic approximation,
counted with respect to the energy of the $1^\textrm{st}$ columnar state
and expressed in units of $J$. The inset is a plot of the ratio of the
energy of the $3^\textrm{rd}$ columnar state over that of the
$4^\textrm{th}$ columnar state.} \label{fig:Comparison of quantum
fluctuations}
\end{figure}

Upon increasing $\Gamma/J$, the classical states remain locally stable
until soft-modes appear in the spin-wave dispersion. For all
columnar states without domain walls of the second type
this takes place at  $\Gamma/J\approx 2.004$, and
for the $3^{\textrm{rd}}$ columnar state at $\Gamma/J\approx 2.373$.
To summarize, harmonic fluctuations partially lift the degeneracy of the
classical ground state manifold in favor of the columnar states having
only domain walls of the $1^{\textrm{st}}$ type.

%*****************************************************************************************************
%*****************************************************************************************************

\section{Plaquette phase}

% \subsection{Instability of the polarized phase}

\subsection{Soft modes and the ground state periodicity}

\begin{figure}[htbp]
\centering
\includegraphics[height=3cm]{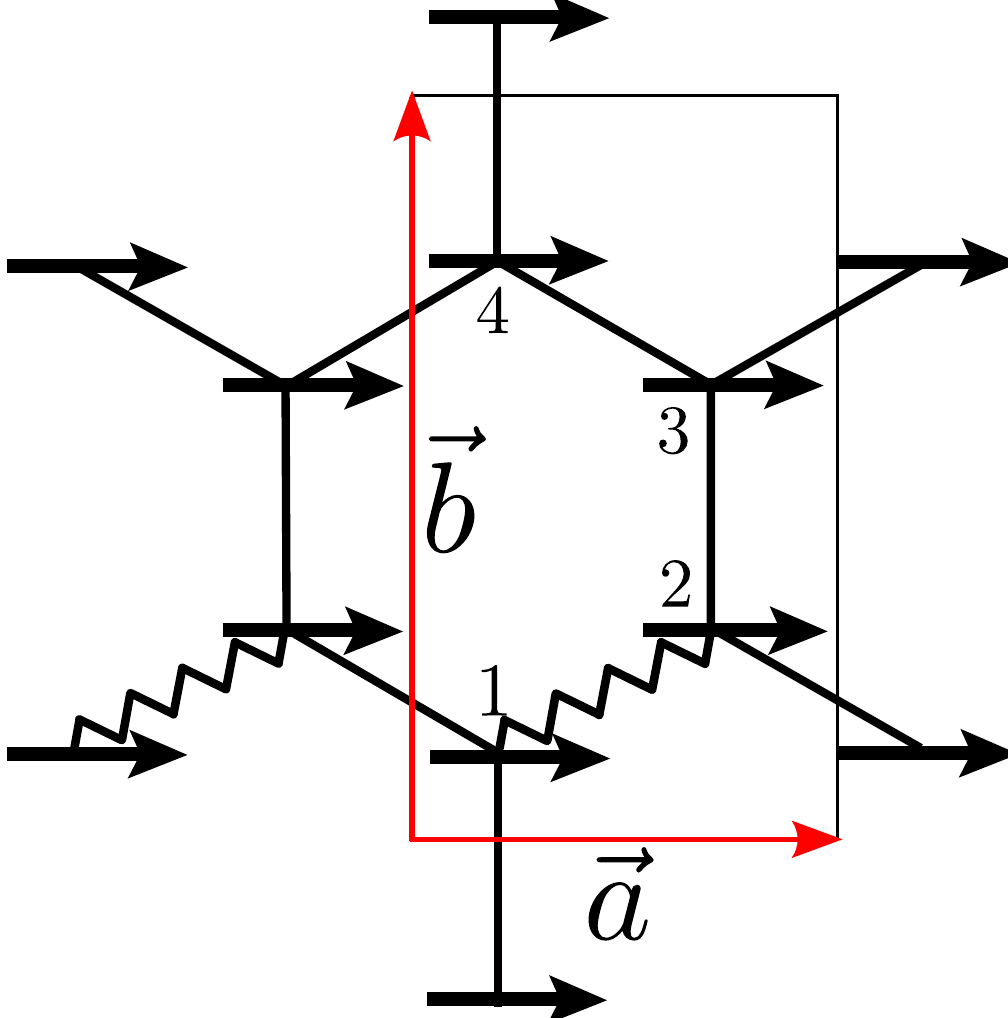}
\caption{(Color online) In the polarized state all spins are aligned along
the magnetic field. The unit cell of this state is the same as {that of
the $1^\textrm{st}$ columnar state: it} is defined by the vectors $\vec{a}$ and
$\vec{b}$.} \label{fig:MaillePolarizedPhase}
\end{figure}

In the limit $J=0$ the Hamiltonian consists simply of a coupling to the
transverse magnetic field $\Gamma$, and the classical ground state is
completely polarized with all spins aligned along the magnetic field in
the $x$ direction. The same state minimizes the classical energy for
sufficiently large ratio $\Gamma/J$. With the choice of gauge of
Fig.~\ref{fig:Choice Of Gauge}, the unit cell of this state contains 4
sites (see Fig.~\ref{fig:MaillePolarizedPhase}).

% Both the Landau-Ginzburg approach of
% Refs.~\onlinecite{moessner1,moessner3} and the linear spin-wave
The analysis of Refs.~\onlinecite{moessner1,moessner3,misguich} indicates
that the polarized phase becomes unstable at
$\Gamma=\Gamma_{c}=\sqrt{6}J$. At this value of the field, soft modes
appear in the dispersion relation at momenta
$(q_x,q_z)=\pm(\frac{\pi}{6|a|},\frac{\pi}{2|b|})$ \textrm{ and }
\makebox{$(q_x,q_z)=\pm(\frac{5\pi}{6|a|},\frac{\pi}{2|b|})$}, triggering
a second order transition to a new phase whose periodicity can be
determined from the $\vec{q}$ points corresponding to the soft modes.

Since any linear combination of these four modes is invariant under
translations by vectors $(3\vec{a}-\vec{b})$ and $4\vec{b}$, a
state associated with them should have the periodicity in real space
imposed by these two vectors that define a unit cell containing $48$ sites
of the honeycomb lattice (Fig.~\ref{fig:48SitePlaquetteSz}). Moreover,
since any linear combination of the four soft modes under the translation
by $2\vec{b}$ just changes sign, this cell should allow a division
into two halves which in the spin representation differ from each other
only by the reflection of all spins about the $x$ axis but in terms of
gauge-invariant variables are identical.

There exists a possibility to
make these two halves really identical in terms of spin representation as
well just by choosing a different gauge shown in Fig.
\ref{fig:PlaquetteSz}. In this gauge a state related to the soft modes
listed above is periodic with a 24-site unit cell defined, for example, by
vectors $3\vec{a}-\vec{b}$ and $2\vec{b}$. However, if one uses
the simplest gauge of Fig.~\ref{fig:Choice Of Gauge} and imposes periodic
boundary conditions along the $x$ and $z$ directions, the periodicity
dictated by the wave vectors of the soft modes requires to use
a cell of size $12\vec{a}\times4\vec{b}$ that contains $192$ sites
of the honeycomb lattice.\cite{moessner3}

\subsection{Numerical minimization of energy}

The minimization of the classical energy using Mathematica minimization
routines for the 192-site system with periodic boundary conditions
have confirmed that the real
periodicity of the classical ground state in the gauge of
Fig.~\ref{fig:Choice Of Gauge} is determined by a 48-site unit cell which
can be divided into two halves in such a way that the second half differs
from the first one by the reflection of all spins about the $x$ axis.
Inside the cell one finds a pattern of six different orientations of the
spins as well as their reflections about the direction of the field.

\begin{figure}[htbp]
 \centering
\subfigure[ Plaquette phase: dimer representation]{\label{fig:Plaquette12x4}
\includegraphics[width=3.5cm]{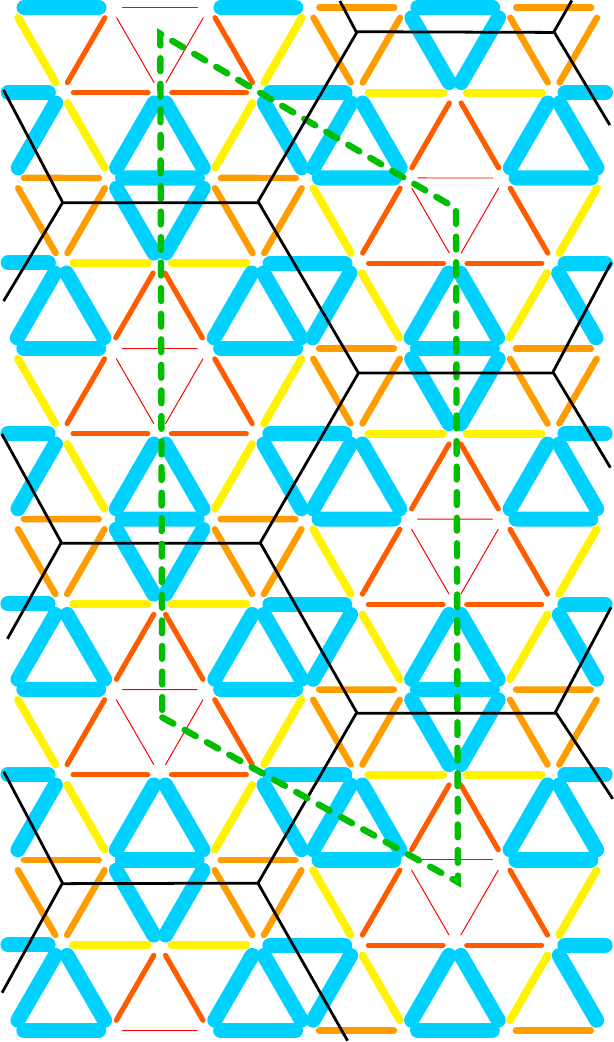}}
\hspace{0.3cm}
\subfigure[ Plaquette phase: unit cell in the spin representation]
{\label{fig:48SitePlaquetteSz}\includegraphics[width=3.5cm]{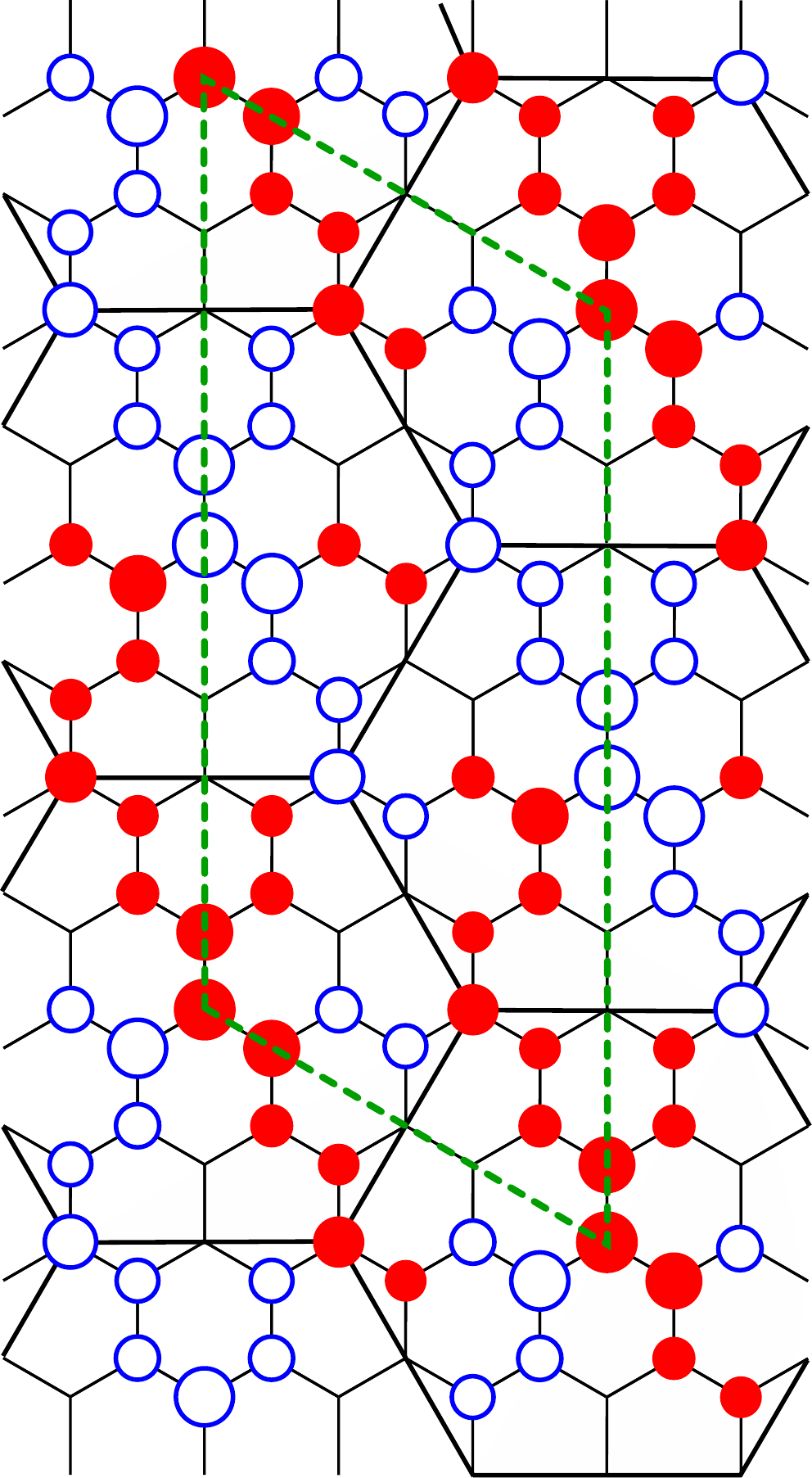}}\\
\subfigure[ Plaquette phase: the smallest unit cell in the spin
representation]
{\label{fig:PlaquetteSz}\includegraphics[height=4.5cm]{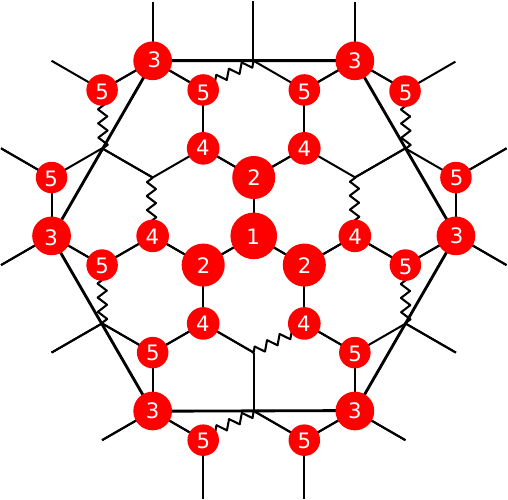}}
\caption{(Color online) (a) The dimer density $d_{ij}$ in the plaquette
phase at $\Gamma/J=2$. The thick blue bonds corresponding to the highest
density ($d_{ij}=\frac{1}{2}$) are organized into 4-site rhombic
plaquettes. On all other bonds {the dimer density satisfies}
$0<d_{ij}<\frac{1}{2}$, the thickness of the bonds
being proportional to $d_{ij}$. \\
%The dashed green rectangle corresponds to the $48$-site unit cell of
% the spin representation.
(b) The spin configuration in the same state in the gauge of
Fig.~\ref{fig:Choice Of Gauge}. The radii of
the circles are proportional to $|S^z_i|$, while
positive and negative values of $S^z_i$ are
represented as full and empty circles.
The green dashed rectangle shows the $48$-site unit cell
$(3\vec{a}-\vec{b})\times 4\vec{b}$. It  can be split into two
halves which differ from each other by {the sign} of $S^z$. \\
(c) The same spin configuration in the gauge that leads to a 24-site unit
cell (large hexagon).
As before antiferromagnetic bonds fixing the gauge %(with $M_{ij}=-1$)
are depicted as {zigzag} bonds. The sites at which the classical spins
have the same values of $S^z_i$ are labelled with the same number. Note
the existence of $6$ sites with $S^z_i=0$.} \label{fig:PlaquettePhase}
\end{figure}

The structure of the state minimizing the classical energy is shown in
figure~\ref{fig:48SitePlaquetteSz}. The radii of the circles are
proportional to the absolute value of the $z$ component of the spins
$|S^z|$ and the different signs of $S^z$ are kept track of by plotting
full and empty circles. $S^x$ is not plotted but is always positive since
the spins tend to align with the magnetic field. The size of the
elementary cell can be reduced to 24 sites by choosing the gauge depicted
in Fig.~\ref{fig:PlaquetteSz} by zigzagged bonds. In this gauge the sign of
$S^z$ is the same for all spins and the spin pattern is centered on
one of the sites of the honeycomb lattice.

Naturally, it is even more convenient  to discuss the structure of an
ordered state in terms of gauge-invariant dimer density $d_{ij}$ defined
by Eq.~(\ref{eq:ClassicalAverageDimerDensity}). In the polarized phase (at
$\Gamma/J>\sqrt{6}$), $S_i^z=0$ for all sites $i$, so that the dimer
density is uniform and equal to $\frac{1}{2}$ on all bonds. Below the
critical magnetic field, ${\Gamma_c}=\sqrt{6}J$, the dimer
density on many bonds becomes smaller than $\frac{1}{2}$.
% shows small deviations about $\frac{1}{2}$, and following
% Ref.~\onlinecite{misguich}, we concentrate on the relative dimer density
% \begin{equation}                         \label{eq:relative dimer density}
% D_{ij}= d_{ij}-\frac{1}{2} =-M_{ij}\frac{S_i^zS_j^z}{2S^2} \leq 0\,.
%\end{equation}
For the pattern of $d_{ij}$ the two halves of the 48-site elementary cell
are identical because the dimer pattern is conserved when reversing the
sign of $S^z$ for all spins. Accordingly, the elementary cell corresponds
to 24 sites of the honeycomb lattice or to 12 sites of the triangular
lattice dual to it. In other terms, the periodicity of the dimer density
pattern is the same as in the $\sqrt{12}\times \sqrt{12}$ phase found
around $V/t=0$ in the QDM on the triangular lattice.
\cite{moessner2,ralko1,ralko2,ralko3}

In Fig.~\ref{fig:Plaquette12x4} the elementary cells
are represented by the large hexagons. Since inside an elementary cell the
dimer density plot displays a pattern of four-site plaquettes
having the highest dimer density (see Fig.~\ref{fig:Plaquette12x4}),
following the convention adopted in the QDM literature
\cite{moessner-raman} we
refer to this phase as the \textit{plaquette phase}. This phase is the
analog of the $\sqrt{12}\times \sqrt{12}$ phase found around $V/t=0$ in
the QDM.\cite{moessner2}

Note that the dimer density plot obtained below
\makebox{$\Gamma/J=\sqrt{6}$} in our calculation
(Fig.~\ref{fig:Plaquette12x4}) differs significantly from the one
presented in Ref.~\onlinecite{misguich}. The two plots have the same
symmetry, $P31m$, but the pattern of Ref.~\onlinecite{misguich} does not
reveal four-site plaquettes. In fact, the difference can be traced back to
the fact that the solution of Ref.~\onlinecite{misguich} was obtained by a
variational calculation in the subspace of linear combinations of the four
soft modes (which minimize the sum of the second and fourth order
contributions to the classical energy), whereas the present solution was
obtained by assuming that the soft modes dictate only its periodicity. The
reason why the two solutions do not have the same asymptotic form when
$\Gamma/J$ tends to $\sqrt{6}$ from below is detailed below in Sec.
\ref{appendix:Difference Plaquette and Misguich Phases} devoted
to the analytical investigation of the plaquette state structure in the
vicinity of the phase transition.

The degeneracy of the plaquette phase is equal to 48 in terms of the spin
representation and to 24 in terms of the dimer representation. Each of the
24 equivalent dimer patterns [one of which is shown in Fig.
\ref{fig:Plaquette12x4}] corresponds to two spin configurations which can
be transformed one into the other by changing the sign of $S^z_i$ for all
spins.

The local stability of the plaquette phase with respect to quantum
fluctuations has been investigated within the gauge of
Fig.~\ref{fig:PlaquetteSz} to reduce the hermitian matrix of the quadratic
bosonic Hamiltonian to a $48\times48$ matrix. The plaquette phase has been
found to be stable in the domain $1.64<\frac{\Gamma}{J}<\sqrt{6}$ with
soft modes appearing at $\vec{q}=0$ when $\frac{\Gamma}{J}\approx 1.64$.

$~$ %*********************************************************************

\subsection{Analytical study of the critical region below $\Gamma_c$
\label{appendix:Difference Plaquette and Misguich Phases}}

In this subsection it will be convenient instead of Eq.~(\ref{eq:Classical
Spins}) to use a {different parametrization of the} classical spins of
norm $S$,
\begin{equation}                    \label{Classical spins - 2}
    \begin{array}{l}
    S_i^x=S\sqrt{1-\rho_i^2} \,,\\ % [1mm]
    S_i^z=S\rho_i \,.
\end{array}
\end{equation}
In the asymptotic regime where the transverse field $\Gamma$ dominates
over nearest-neighbor interactions, we are in the polarized phase with
$S^z_i=0$ ($\rho_i=0)$. Upon decreasing the transverse field the components
$S^z_i$ are expected to deviate from zero. To sixth order
in the $\rho_i$'s, the classical energy of the model is given by
\begin{equation}\label{eq:Sixth order energy}
E=-J\sum_{\langle i,j\rangle}{M_{i,j}\rho_i\rho_j}
    -\Gamma\sum_{i}{\left(1-\frac{\rho_i^2}{2}-\frac{\rho_i^4}{8}-\frac{\rho_i^6}{16}-\ldots\right)}\,,
\end{equation}
Let us denote by
$\rho_{\vec{R}_i,n}=\sum_{\vec q}\rho_{\vec{q},n}e^{i\vec{R}_i\vec{q}}$
with $n=1,\dots,4$ the values of $\rho_i$ on the four sublattices (see
Fig.~\ref{fig:MaillePolarizedPhase}). Since $\rho_{\vec{R}_i,n}$ is real,
${\rho}_{\vec{q},n} = {\rho}^*_{-\vec{q},n}$.
The energy per site ${\cal E}$ % (\ref{eq:Sixth order energy})
is then given by
\begin{eqnarray}\label{eq:Energy to Sixth order in rho after FT}
%\begin{array}{ll}
{\cal E}&=&{\cal E}_{J=0}- \frac{J}{8}\displaystyle{\sum_{n,n',\vec{q}}{{\rho}_{-\vec{q},n}\left[\hat{M}(\vec{q})-\frac{\Gamma}{J}\hat{\id}\right]_{n,n'}{\rho}_{\vec{q},n'}}}\nonumber \\
       & &+\frac{\Gamma}{32}\displaystyle{\sum_{\substack{n,\vec{q}_1,\vec{q}_2, \\ \vec{q}_3,\vec{q}_4}}
                                        {\left(\prod_{i=1}^4\rho_{\vec{q}_i,n}\right)\delta_{\vec{q}_1+\vec{q}_2+\vec{q}_3+\vec{q}_4,\vec{G}}}}\\
       & &+\frac{\Gamma}{64}\displaystyle{\sum_{\substack{n,\vec{q}_1,\vec{q}_2,\vec{q}_3, \\ \vec{q}_4,\vec{q}_5,\vec{q}_6}}
                                        {\left(\prod_{i=1}^6\rho_{\vec{q}_i,n}\right)\delta_{\vec{q}_1+\vec{q}_2+\vec{q}_3+\vec{q}_4+\vec{q}_5+\vec{q}_6,\vec{G}}}}+\ldots\,,\nonumber
%\end{array}
\end{eqnarray}
where % $N$ is the total number of sites,
$\vec{G}$ is a vector belonging to
the reciprocal lattice of the lattice defined by the vectors $\vec{a}$ and
$\vec{b}$, and $\hat{M}({\vec{q}})=$
\[
    \left(\begin{array}{cccc}
                 0          & -1+e^{-iq_x|a|}   &      0            & e^{-iq_z|b|} \\
            -1+e^{iq_x|a|}  &         0         &      1            & 0 \\
                     0      &         1         &      0            & 1+e^{iq_x|a|} \\
            e^{iq_z|b|}     &         0         &   1+e^{-iq_x|a|}  & 0
    \end{array}\right)\,
\]
is the Fourier transform of the interaction matrix. The analysis of the
second-order terms in (\ref{eq:Energy to Sixth order in rho after FT})
shows \cite{moessner1,moessner3} that the paramagnetic solution $\rho_i=0$
becomes unstable at ${\Gamma}/{J}=\sqrt{6}$ at the wavevectors
$\vec{q}_A=\left(\frac{\pi}{6|\vec{a}|},\frac{\pi}{2|\vec{b}|}\right)$,
$\vec{q}_B=\left(\frac{5\pi}{6|\vec{a}|},\frac{\pi}{2|\vec{b}|}\right)$,
$-\vec{q}_A$, and $-\vec{q}_B$, indicating a transition to a phase of
periodicity $(3\vec{a}-\vec{b})\times 4\vec{b}$.

The approach of Ref.~\onlinecite{misguich} consists in keeping in the
energy functional (\ref{eq:Energy to Sixth order in rho
after FT}) % the associated to
only the critical modes with $\vec{q}=\pm \vec{q}_A$ and $\vec{q}=\pm
\vec{q}_B$ whose amplitudes are described by Fourier coefficients
\begin{equation}\label{eq:Non zero Fourier coefficients}
 \begin{array}{l}
\vec{\rho}_{\vec{q}_A}=|\rho_{A}|e^{i\phi_{A}}\vec{u}_A\,,~~~ %   \\ [3mm]
\vec{\rho}_{\vec{q}_B}=|\rho_{B}|e^{i\phi_{B}} \vec{u}_B\,,
\end{array}
\end{equation}
where
\begin{equation}\label{eq:eigenvectors}
\begin{array}{rcl}\vec{u}_A & = &
\left(1,e^{i\frac{7\pi}{12}},Fe^{i\frac{7\pi}{12}},Fe^{-i\frac{3\pi}{2}}\right) \\
\vec{u}_B & = &
\left(F,Fe^{i\frac{11\pi}{12}},{e^{i\frac{11\pi}{12}}},{e^{-i\frac{3\pi}{2}}}\right)
\end{array}
\end{equation}
are the eigenvectors of $\hat M(\vec{q}_A)$ and $\hat M(\vec{q}_B)$
associated to the eigenvalue $\sqrt{6}$ and
\begin{equation}\label{eq:F}
    F=2\sin\frac{5\pi}{12}=\frac{1+\sqrt{3}}{\sqrt{2}}\;.
\end{equation}
%\begin{equation}\label{eq:definition of u and v}
% \begin{array}{l}
% M(q_A)\vec{u}_A=\sqrt{6}\vec{u}_A \\[3mm]
% M(q_B)\vec{u}_B=\sqrt{6}\vec{u}_B \\
% \end{array}
%\end{equation}

In the framework of this approach, ${\cal E}_0^{(4)}$,
the sum of the second and fourth order contributions to % $\cal E$,
Eq.~(\ref{eq:Energy to Sixth order in rho after FT}),
is given by:
\begin{equation}\label{eq:E4 qo modes}
 \begin{array}{lll}
{\cal E}_0^{(4)}&=&-\frac{1}{2}(\Gamma_c-\Gamma){(1+F^2)}{\left[|\rho_{A}|^2+|\rho_{B}|^2\right]}
\\ [4mm]
                   & & +\frac{3\Gamma}{2} F^2 \left[ |\rho_{A}|^2+|\rho_{B}|^2\right]^2
\end{array}
\end{equation}
and  depends only on $|\rho_{A}|^2+|\rho_{B}|^2$.  \cite{moessner1,moessner3}

The minimum of ${\cal E}_0^{(4)}$ is achieved when
\begin{equation}\label{eq:Behaviour rhoA and rhoB}
 |\rho_{A}|^2+|\rho_{B}|^2=\frac{1+F^2}{6F^2}\frac{\Gamma_c-\Gamma}{\Gamma}\,,
\end{equation}
from which it follows that, to leading order,
\makebox{$|\rho_{A}|\sim|\rho_{B}|\sim (\Gamma_c-\Gamma)^{\frac{1}{2}}$}
and ${\cal E}_0^{(4)}\sim (\Gamma_c-\Gamma)^2$. However,
condition (\ref{eq:Behaviour rhoA and rhoB}) leaves both the ratio
$|\rho_B|/|\rho_A|$ and the phases $\phi_A$ and $\phi_B$ completely
undefined. To find them one has to consider also the sixth order term in
Eq.~(\ref{eq:Energy to Sixth order in rho after FT}),
\cite{moessner3,misguich} which for the critical modes reduces to
\begin{equation}\label{eq:E6 qo modes}
\begin{array}{lll}
{\cal E}_0^{(6)}&=&\frac{5\Gamma}{8}(1+F^6)\left[|\rho_{A}|^2+|\rho_{B}|^2\right]^3
\\ [3mm]
         &+ &\frac{3\Gamma}{2} F^3\left[|\rho_{A}|^5|\rho_{B}|\cos(5\phi_{A}-\phi_{B})\right.\\ [3mm]
         & & \hspace*{8mm}\left.+|\rho_{B}|^5|\rho_{A}|\cos(5\phi_{B}-\phi_{A})\right]\,.
\end{array}
\end{equation}
The general structure of Eq.~(\ref{eq:E6 qo modes}) has been derived in
Ref. \onlinecite{moessner3} from the symmetries of the problem.

According to the previous discussion, to leading order,
${\cal E}_0^{(6)}\sim(\Gamma_c-\Gamma)^3$. For all values of the
amplitudes $|\rho_A|$ and $|\rho_B|$, the expression in the right-hand
side of Eq.~(\ref{eq:E6 qo modes}) is minimal when both cosines are equal
to $-1$. This selects the phases:
\begin{equation}\label{eq:phiA and phiB no q1 modes}
    \begin{array}{lcr}
        \phi_{A}=\frac{\pi}{6}+\frac{\pi}{12}p\,, & & % -\frac{\pi}{24}(1+2n_1)+\frac{5\pi}{24}(1+2n_2) (equivalent to choice n_1=-m + 5n_2)
        \phi_{B}=-\frac{\pi}{6}+\frac{5\pi}{12}p\,, %-\frac{5\pi}{24}(1+2n_1)+\frac{\pi}{24}(1+2n_2) (equivalent to choice n_1=-m + 5n_2)
    \end{array}
\end{equation}
where $p$ is an integer, yielding $24$ independent sets $(\phi_A,\phi_B)$.
The variation of ${\cal E}_0^{(6)}$ with respect to $|\rho_A|$ and $|\rho_B|$
under the constraints (\ref{eq:Behaviour rhoA and rhoB}) and (\ref{eq:phiA
and phiB no q1 modes}) then selects either $|\rho_B|/|\rho_A|=F$ or
$|\rho_B|/|\rho_A|=F^{-1}$.  All $48$ solutions thus generated
correspond to the same dimer pattern (shifted or/and rotated) found in
Ref.~\onlinecite{misguich}
and thus we recover the $48$ fold degeneracy discussed in Ref.~\onlinecite{moessner3}.

The approach described above is based on the assumption that all other
modes would only contribute to the energy expansion to higher order. We
shall now show that, since when considering only the critical modes
% with $\vec{q}=\pm \vec{q}_A$ and $\vec{q}=\pm \vec{q}_B$
one has to push the expansion to order 6, this assumption is not valid
because some second- and fourth-order terms involving noncritical modes
also make contributions of order $(\Gamma_c-\Gamma)^3$ which are essential
for determining $\phi_A$ and $\phi_B$.

The dominant terms coupling the critical modes with $\vec{q}=\pm
\vec{q}_A$ and $\vec{q}=\pm \vec{q}_B$ with extra modes are expected to be
linear in the amplitudes of these extra modes and of the third order in
the amplitudes of critical modes. The conservation of the total momentum
then imposes on the wavevectors of these extra modes the condition:
\begin{equation}
 \vec{q}=m_A \vec{q}_A + m_B \vec{q}_B\,,
\end{equation}
where $m_A$ and $m_B$ are integers and $m_A+m_B$ is odd. In the first
Brillouin zone there are only two wavevectors compatible with this
condition: $\vec{q}_C=2\vec{q}_A-\vec{q}_B$ and $-\vec{q}_C$. Let us
denote the Fourier coefficients associated to the modes with
$\vec{q}=\vec{q}_C$ by
$\overline{\rho}_{n}=|\overline{\rho}_{n}|e^{i\overline{\phi}_n}$, where
$n=1,\ldots,4$ refers to the number of the sublattice. The terms in
the energy functional which are linear and harmonic in $\overline{\rho}_n$ are
\begin{eqnarray} \label{eq:E 4 q1 modes}
{\cal E}_1^{(4)}& =& -\frac{J}{4}\displaystyle\sum_{n,n^\prime}\overline{\rho}_n^* \left[\hat{M}(\vec{q}_C)-\frac{\Gamma}{J}\hat{\id}\right]_{n,n^\prime}\overline{\rho}_{n^\prime}\nonumber \\
                    &+&\frac{\Gamma}{8}\sum_{n=1}^4\left(R_n{\bar \rho}_n+\mbox{c.c.}\right)\,,
\end{eqnarray}
with
\begin{equation} \label{eq: R_n}
\begin{array}{ccc}
R_n&=&\rho_A^3{(u_A)}_n^3+3(\rho_A^{*})^2({u_A^*})_n^2\rho_B{(u_B)}_n \\[3mm]
   & &+3\rho_A{(u_A)}_n(\rho_B^{*})^2({u_B}^*)_n^2+\rho_B^3 {(u_B)}_n^3\,.
\end{array}
\end{equation}

\noindent The variation of Eq.~(\ref{eq:E 4 q1 modes}) with respect to
$\overline{\rho}_{n}^*$ gives
\begin{equation} \label{eq: rho_n}
\bar
\rho_n=\frac{\Gamma}{2J}\sum_{n'}\left[\hat{M}(\vec{q}_C)-\frac{\Gamma}{J}
\hat{\id} \right]^{-1}_{nn'}R_{n'}^*\,.
\end{equation}
Injecting Eq.~(\ref{eq: rho_n}) into Eq.~(\ref{eq:E 4 q1 modes}) we obtain
\begin{eqnarray} \label{eq: E 4 q1 modes after simplification}
%\begin{array}{lll}
{\cal E}_1^{(4)}  &=& -\Gamma h\left({\Gamma}/{J}\right)\left[|\rho_A|^2
                      +{|\rho_B|^2}\right]^3 \\%[3mm]
& & -\Gamma g\left({\Gamma}/{J}\right)\left[|\rho_A|^5|\rho_B|
    \cos(5\phi_A-\phi_B)\right. \nonumber\\%[3mm]
& & \hspace*{18mm}
+\left.|\rho_A||\rho_B|^5\cos(5\phi_B-\phi_A)\right]\,,\nonumber
%\end{array}
\end{eqnarray}
where we have introduced the notation
\begin{eqnarray} \nonumber
 h(\gamma)&=&\frac{\gamma}{8(\gamma^2-3)}\left\{ \gamma(1+F^6)+6\sqrt{2}(3F^2-1)\right\}\,, \\
 g(\gamma)&=&\frac{3\gamma F}{4(\gamma^2-3)}\left\{ 4\gamma F^2+3\sqrt{2}(2F^2-1)\right\}\,.
\nonumber
\end{eqnarray}
Eq.~(\ref{eq: rho_n}) proves that $\bar \rho_{n}$ scales as
\begin{equation}\label{eq:Behaviour q1 modes}
|\bar \rho_{n}|
\sim|\rho_A|^3\sim|\rho_B|^3\sim(\Gamma_c-\Gamma)^\frac{3}{2}\,,
\end{equation}
leading to ${\cal E}_1^{(4)}\sim(\Gamma_c-\Gamma)^3$. So it is clear that this
contribution cannot be neglected since it is of the same order as
${\cal E}_0^{(6)}$, and that other contributions involving non-critical modes
such as e.g. sixth order terms will be of higher order. This means that
the phases of the critical modes have to be determined by minimizing the
sum of ${\cal E}_0^{(6)}$ and ${\cal E}_1^{(4)}$.
The contribution to this expression depending on the phases reads:
\begin{eqnarray}\label{eq: Energy dependance on the phases}
\nonumber -\Gamma\left[g\left(\frac{\Gamma}{J}\right)-{\frac{3}{2}}F^3
\right] \left[|\rho_A|^5|\rho_B|\cos(5\phi_A-\phi_B) +\right. && \\ \left.
+|\rho_A||\rho_B|^5\cos(5\phi_B-\phi_A)\right]\,. &~&~ \nonumber
\end{eqnarray}
Now $g({\Gamma}/{J})-(3/2)F^3$ is positive for $\Gamma/J>\sqrt{3}$. Therefore,
since we are interested in the domain just below \makebox{$\Gamma/J =
\sqrt{6}$}, the energy is minimal when both cosines are equal to $+1$.
This selects the phases:
\begin{equation}\label{eq:phiA and phiB with q1 modes}
        \phi_{A}=\frac{\pi}{12}p\,,~~~ %& \quad & %\frac{\pi}{12}(5n_2-n_1)
%       (equivalent for the choice n_1=-m + 5n_2)\\ [2mm]
        \phi_{B}=\frac{5\pi}{12}{p}\,,        %\frac{\pi}{12}(n_2-5n_1)
\end{equation}
where $p$ is an integer. This leads again to $24$ independent sets
$(\phi_A,\phi_B)$. In addition, minimizing ${\cal E}_0^{(6)}+{\cal E}_1^{(4)}$
with respect to the amplitudes $|\rho_A|$ and $|\rho_B|$ under the constraint
(\ref{eq:Behaviour rhoA and rhoB}) selects, as before, either
$|\rho_B|/|\rho_A|=F$ or $|\rho_B|/|\rho_A|=F^{-1}$. The 48 resulting
solutions correspond to the 24 equivalent dimer patterns which can be obtained
from the one shown in Fig.~\ref{fig:Plaquette12x4}.
% by shifting or/and rotating it.}
The difference between
Eq.~(\ref{eq:phiA and phiB no q1 modes}) and Eq.~(\ref{eq:phiA and phiB
with q1 modes}) explains the qualitative difference between the structures
of the plaquette phase found in this work and the solution of
Ref.~\onlinecite{misguich}, which does not disappear even when the
amplitudes of the $\vec{q}=\pm\vec{q}_C$ modes become negligible as
compared to those of the critical modes.

$~$%**********************************************************************

\section{Intermediate mixed phases}

\begin{figure}[bthp]
\centering
\subfigure[ Mixed phase: dimer representation]{\label{fig:MixedPhase12x4}
\includegraphics[height=4.5cm]{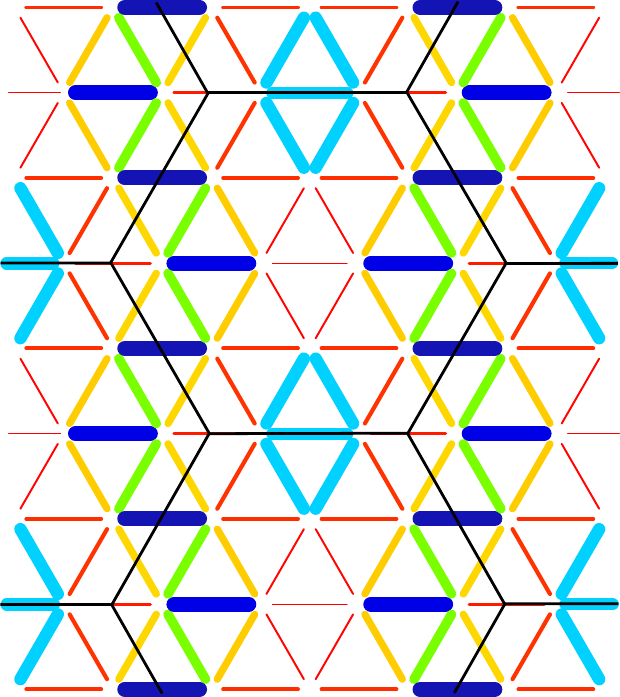}}\\
\subfigure[ Mixed phase: spin representation]{\label{fig:MixedPhaseSz}
\includegraphics[height=4.5cm]{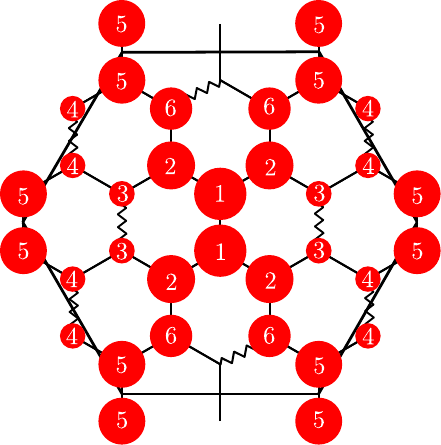}}
\caption{(Color online) (a) The dimer density $d_{ij}$ in the mixed phase at $\Gamma/J=1.72$.
 The thickness of the bonds is proportional to  $d_{ij}$.
 The dimer densities are also emphasized by the colors of the bonds
 ranging from red (the lowest densities) to dark blue (the highest densities,
 $d_{ij}>\frac{1}{2}$).
 The bonds with $d_{ij}\geq\frac{1}{2}$ are organized in an alternating
 pattern of plaquettes and columns.
 (b) The spin configuration in the same state in the gauge that leads
 to a 24-site unit cell (large hexagon).
% Same as Fig.~\ref{fig:PlaquetteSz} for the mixed phase.
 Note the existence of $2$ sites at which $S^z_i=0$.}
\label{fig:MixedPhase}
\end{figure}

During the numerical minimization of the classical energy
for the 192-site system with periodic boundary conditions %in the $x$ and $z$ direction
an additional intermediate phase was found to exist between the columnar
and the plaquette phases.
%(some details on how {we came across this} intermediate phase are presented in Appendix \ref{appendix:Gamma Gamma4model}).
We refer to this intermediate phase as the mixed phase because in
the dimer representation the bonds with larger dimer densities,
$d_{ij}\geq\frac{1}{2}$, are arranged in an alternating pattern of
plaquettes and columns (see Fig.~\ref{fig:MixedPhase12x4}).
%{\cb The integer $n=1$ indicates that the plaquette structures are separated by one column}
The mixed and plaquette phases have the same translational symmetries. However, the
point group symmetries of the gauge-invariant dimer patterns in the two
phases are different: $P31m$ for the plaquette phase (see Figs.
\ref{fig:Plaquette12x4}) and $Cmm$ for the mixed phase (see
Fig.~\ref{fig:MixedPhase12x4}). The phase transition between these two
phases has to be of the first order, since the symmetry groups are not
such that one is a subgroup of the other.

As in the case of the plaquette phase, the size of a unit cell of the
mixed phase can be reduced from $48$ sites for the standard gauge shown in
Fig.~\ref{fig:Choice Of Gauge} to $24$ sites in the gauge of
Fig.~\ref{fig:PlaquetteSz}, see Fig.~\ref{fig:MixedPhaseSz}. In this gauge
the spin pattern consists of spins with the same sign of $S^z_i$
having seven different orientations, one of which is in the direction of the field.
In contrast to the spin pattern in the plaquette phase, which is centered
on one of the sites of the honeycomb lattice (Fig.~\ref{fig:PlaquetteSz}),
in the mixed phase this pattern is centered on one of the bonds of the lattice
(Fig.~\ref{fig:MixedPhaseSz}), which explains the difference in symmetry
between the two states.
\noindent
The degeneracy of the mixed phase is equal to 36 in terms of the dimer
representation and to 72 in terms of the spin representation. Each of the
36 equivalent dimer patterns
% [one of which is shown in Fig. \ref{fig:MixedPhase12x4}]
corresponds to two spin configurations which can be transformed one into
the other by changing the sign of $S^z_i$ for all spins.
The stability of the mixed state with respect to small fluctuations has been
investigated with LSWT in the gauge producing a 24-site unit cell, and
this phase has been found to be stable in the range
$1.394\lesssim{\Gamma}/{J}\lesssim 1.774$.

\begin{figure}[bthp]
 \centering
 \includegraphics[width=8cm]{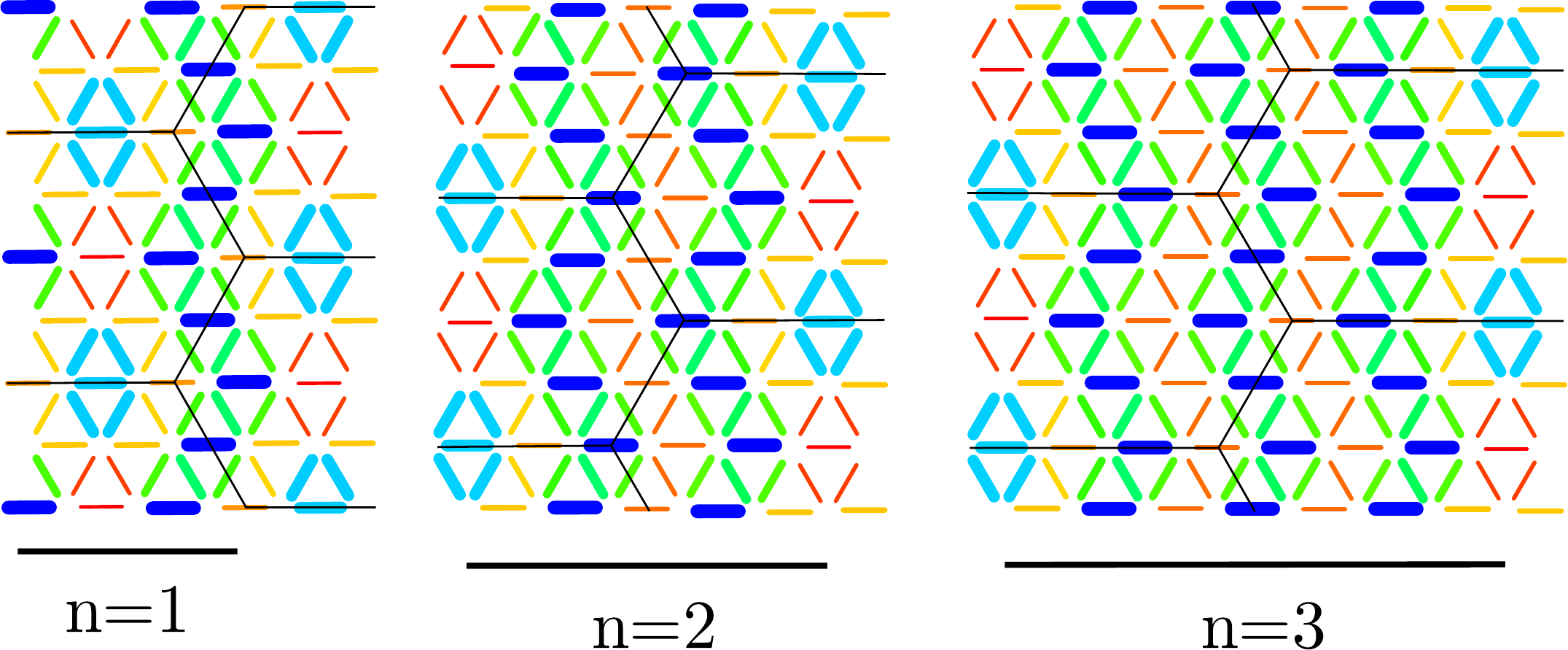}
 \caption{ (Color online) Dimer patterns %  density $d_{ij}$ of
 in the mixed states with $n\leq3$, where $n$ denotes the number of
 columns separating the plaquette patterns. The notation is the same as in Figs.
 \ref{fig:Plaquette12x4} and \ref{fig:MixedPhase12x4}.
 % The thickness of the bonds is proportional to $d_{ij}$.
 % The dimer densities are also emphasized by the colors of the
 % bonds ranging from red (the lowest densities) to dark blue
 % (the highest densities, $d_{ij}>\frac{1}{2}$).
 The bonds with $d_{ij}\geq\frac{1}{2}$ are organized in an alternating
 pattern of plaquettes and columns. Thin lines show the boundaries between unit cells.
 }
\label{fig:MixedPhases}
\end{figure}

The existence of the mixed state whose structure is shown in Fig.~\ref{fig:MixedPhase}
suggests that there can also exist states in which the
straight rows of plaquettes are still equidistant but separated not by single columns
but by a larger number of columns which below is denoted by $n$ (see Fig.~\ref{fig:MixedPhases}).
From now on we number such mixed states by the index $n$ and call
the simplest mixed state discussed above the first mixed state.

It is not hard to understand that
the unit cell of the second mixed state (in the optimal gauge in which
the sign of $S^z$ is the same for all spins) has exactly the same symmetry
as the unit cell of the first mixed state and can be obtained from it by
adding on each side eight more sites. The successive repetition of this
procedure allows one to construct the unit cell for any integer $n$ and to
find that it contains $8(2n+1)$ sites.
However, due to the symmetry of the unit cell, the number of non-equivalent sites
only increases by four when $n$ increases by one, which leads
to $4n+3$ non-equivalent sites.

% The general expression of the classical energy per site of a mixed $n$ state in terms of $2(2n+1)$ independant
% variables (the sites corresponding to the plaquette structre being fixed at $S^z=0$) is presented in Eq.~(\ref{eq:Classical energy mixed states}).
%
% \begin{widetext}
% \begin{eqnarray} \label{eq:Classical energy mixed states}
% \nonumber
% E_n & = & -\frac{J}{8(2n+1)}[\cos^2\theta_1+4\cos\theta_1\cos\theta_2 +2(-1)^n\cos^2\theta_{2n+2} +\sum_{k=1}^n(4\cos\theta_{4k-2}\cos\theta_{4k-1}-2\cos^2\theta_{4k-1} +4\cos\theta_{4k-1}\cos\theta_{4k} \\
% \nonumber
%     &   & \hspace*{5mm} + 4\cos\theta_{4k}\cos\theta_{4k+1} +2\cos^2\theta_{4k+1}+4\cos\theta_{4k+1}\cos\theta_{4k+2}) - 4\sum_{k=1}^n{(-1)^k\cos\theta_{2k}\cos\theta_{4n+4-2k}}] \\
%     & - & \frac{\Gamma}{8(2n+1)}\left(2+2\sin\theta_1+4\sum_{k=2}^{4n+2}\sin\theta_k\right)
% \end{eqnarray}
% \end{widetext}
%

For $n\leq 7$ we have performed a numerical minimization of the energy for the unit cells
corresponding to such structures, and we have found that, upon decreasing $\Gamma/J$, the energy
of the second mixed state first becomes lower than that of the first mixed state, after
what the energy of the third mixed state becomes lower than that
of the second mixed state, and so on. Tab. I % ~\ref{Tab:Mixed transitions}
summarizes the values of $\Gamma/J$ at which the
transition between the $n$th and $(n+1)$th mixed states takes place and reports
the width of the region in which the $n$th mixed state has the lowest energy.
It can be seen that for $n>1$ this width is scaled down
by a factor of the order of $50$ each time $n$ increases by 1.
This means that $\Gamma_c^{n,n+1}$ approaches a finite limit exponentially fast.
The extrapolation shows that the accumulation point of $\Gamma_c^{n,n+1}$
at $n\rightarrow \infty$ is $\Gamma_c^\infty/J= 1.67612786261$.
Below this field columnar states have the lowest classical energy.

\begin{table}[htbp] \label{Tab:Mixed transitions}
\begin{centering}
\begin{tabular}{|c|c|c|}
\hline
% \multicolumn{3} {|c|}{Transitions between mixed states} \\ \hline
$n$ & $\Gamma_c^{n,n+1}/J$              &   $\Delta\Gamma^{n}/J$  \\ \hline
 $1$      &  \multirow{2}{*}{$1.69372479498$} &   $4.3\times 10^{-2}$              \\[2mm]
 $2$      &  \multirow{2}{*}{$1.67655449242$} &   $1.7\times 10^{-2}$         \\[2mm]
% Mixed $(n_1,n_2)=(2,3)$ & \multirow{2}{*}{$1.67655449053$}\\[2mm]
 $3$      &  \multirow{2}{*}{$1.67613666486$} &   $4.2\times 10^{-4}$         \\[2mm]
 $4$      &  \multirow{2}{*}{$1.67612802659$} &   $8.6\times 10^{-6}$         \\[2mm]
 $5$      &  \multirow{2}{*}{$1.67612786551$} &   $1.6\times 10^{-7}$         \\[2mm]
 $6$      &  \multirow{2}{*}{$1.67612786267$} &   $2.8\times 10^{-9}$         \\[2mm]
 $7$      &      $\hdots$         &  $\hdots  $                                \\[3mm] \hline
\end{tabular}
\caption{Critical field $\Gamma_c^{n,n+1}$
of the transition between the $n$th and $(n+1)$th mixed states.
The last column shows $\Delta\Gamma^{n}= \Gamma_c^{n-1,n}-\Gamma_c^{n,n+1}$,
the field range in which the $n$th mixed state
has lower classical energy than the $(n-1)$th and the $(n+1)$th states
($\Gamma_c^{0,1}$ refers to the transition
between the plaquette and the first mixed state).
}
\end{centering}
\end{table}

\noindent Note that it was impossible to discover any of the mixed states
with $n > 1$ during the minimization of the energy for the 192-site cell
(with periodic boundary conditions and the standard gauge of Fig.
\ref{fig:Choice Of Gauge})
which was instrumental in discovering the $n=1$ mixed state.
The reason is very simple - the periodicity of all the states
with $n>1$ is incompatible with the periodic boundary conditions
implemented in this 192-site cell.

The existence of such a sequence of phase transitions suggests that the main contribution
to the energy of the $n$th mixed phase (counted off from the energy of a columnar state)
is proportional to the density of linear defects (vertical rows of plaquettes) whose energy
can be considered as linearly dependent on $\Gamma$, whereas
the main correction to this energy comes from the repulsion of nearest defects,
which decreases exponentially fast with the distance between them.
This was checked at $\Gamma=\Gamma_c^\infty$ where the proper energy of a linear
defect changes sign, and indeed we have found that
the energies of different states are compatible with an interaction
of linear defects that is exponential in the distance between them.
This makes us confident that the narrow region above $\Gamma^\infty$
has to contain an infinite sequence of mixed phases with all integer
indices $n$.

It is well known that in a system consisting of a sequence of linear defects
there can also exist phases with more complex structures, in which the linear
defects are not equidistant. In terms of our problem such phases would
correspond to a regular alternation of, for example, $n$ and $n+1$
columns, or of $n$, $n$ and $n+1$ columns, {\em etc.}, leading to what is known
as a devil's staircase. \cite{Bak} Usually such phases appear
in a phase diagram if the interaction of more distant defects is also repulsive,
whereas when the interaction between next-to-nearest defects is attractive, one
gets a direct transition from the $n$th to the $(n+1)$th phase without the presence of
an intermediate $(n,n+1)$ phase.

We have verified numerically that in our system the energy of the $(1,2)$
mixed state is never lower than either the energy of the first state
or that of the second mixed state, which means that it cannot be present
in the phase diagram. Quite surprisingly, the situation with the $(2,3)$ phase
is different, and in a narrow interval around $\Gamma_c^{2,3}$
[from $\Gamma_c^{2,3}-1.3 \times 10^{-9}$ to $\Gamma_c^{2,3}+1.9 \times 10^{-9}$]
its energy is lower than the energies
of the second and third mixed states. One can estimate that even if some other
complex phases do exist, the field range where any of them minimizes
the energy will be at least a couple of orders of magnitude smaller than the already
extremely narrow interval of the existence of the $(2,3)$ state, so we decided
not to pursue the investigation of this point any further since it cannot
be of much relevance.

A more important question is whether the plaquette and the first mixed states
may be separated by a region where there appear mixed states of a different type,
in which the density of columns is lower than in the first mixed state, so that
the neighboring columns are separated by domains of plaquette state.
Such a scenario seems to us to be impossible however for the following reasons.

The comparison of Fig.~\ref{fig:PlaquetteSz} with Fig.~\ref{fig:MixedPhaseSz}
suggests that the structure of the first mixed state is very close to what one
would obtain by constructing the superposition of two plaquette states centered
on neighboring sites of the lattice (and letting this superposition relax).
Therefore one can interpret these two states
as different manifestations of a unique state which can move around in a complex
periodic potential with minima both at the positions corresponding
to lattice sites and at the positions corresponding to the middles of lattice bonds.
For $\Gamma > \Gamma_c^{0,1} = 1.73690830184 J$,
the minima located at lattice sites are the lowest, whereas
for $\Gamma < \Gamma_c^{0,1}$, the minima located at the middle of lattice bonds
are the lowest. Exactly at $\Gamma = \Gamma_c^{0,1}$ all these minima have equal
depths. This picture can be confirmed by constructing a family of states which
continuously interpolates between the plaquette and the first mixed state,
which allows a numerical analysis of the effective potential discussed above.
This analysis reveals that at $\Gamma = \Gamma_c^{0,1}$ the barrier separating
unequivalent (but equal) minima is very low ($\sim 1.07 \times 10^{-5}J$ per site).
Nonetheless, any attempt to construct
a state which somewhere looks like the plaquette state and elsewhere like the first
mixed state would force the system to overcome this barrier in some places.
This will increase its energy in comparison with that of the plaquette or of
the first mixed state.

The numerical evidence in favor of this conclusion comes from observing that the
state which would differ from the first mixed state by having half its density
of columns has a periodicity which is compatible with the 192-site cell used in
our numerical energy minimization. Therefore, if at $\Gamma=\Gamma_c^{0,1}$ the energy
of this state was lower than that of the plaquette and of the first mixed states,
this state would be accessible during this minimization procedure.
To be on the safe side, we have also performed a minimization of the energy for the cell
whose periodicity in addition to the formation of the plaquette and of the first mixed
states
allows for the appearance of the states which differ from the first mixed state by
keeping only one column out of three (or two out of three), but this
has not allowed us either to find any state with energy lower than that of the plaquette or of
the first mixed state. This gives an additional evidence in favor of our conclusion
that the phase transition between the plaquette and the first mixed states should
be a direct one without any intermediate phases with a more complex structure.

%**************************************************************************
%**************************************************************************

\section{Phase diagram}

\subsection{Classical phase diagram}

The classical phase diagram consists of 4 regions: (i) The columnar phase,
which is highly degenerate since all columnar states have the same energy.
It extends up to $\Gamma/J\approx 1.676$; (ii) The region of mixed states
with columnar patterns separated by straight rows of plaquettes
in the interval $1.676 \lesssim \Gamma/J \lesssim 1.737$;
(iii) The plaquette phase, with a 24-site unit cell, in
the range $1.737 \lesssim \Gamma/J\le \sqrt{6}\approx 2.45$; (iv)
The fully polarized phase with all spins pointing in the direction
of the field for $\Gamma/J>\sqrt{6}$.
The transition from the fully polarized phase to the plaquette
phase is a second-order one, all other transitions being of the first order.
These results are summarized in Fig.~\ref{fig:ClassicalPhaseDiagram}.

\begin{widetext}

\begin{centering}
\begin{figure}[bthp]
\includegraphics[width=16cm]{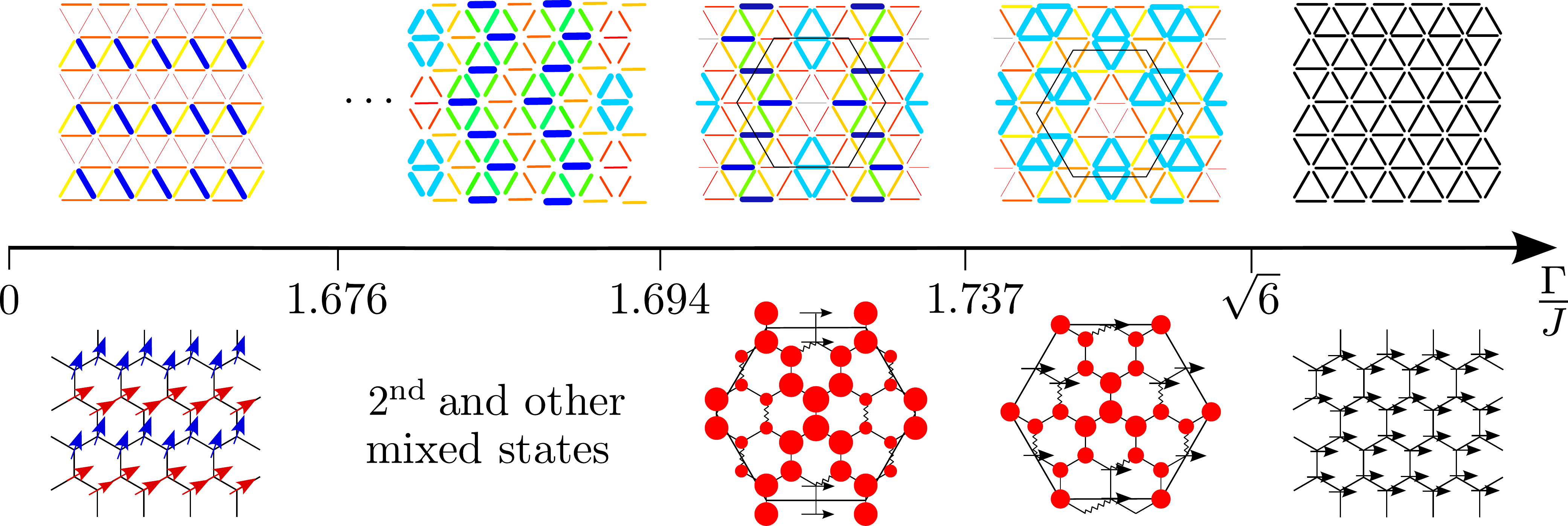}
\caption{(Color online) Classical phase diagram in the dimer language
(above) and in the spin language (below). In the dimer
representation the thickness of the bonds is proportional to the dimer
density. Thick blue bonds correspond to the highest dimer density. In the
spin representation the radii of the circles are proportional to $S_i^z$
and arrows indicate the orientation of the classical spins.}
\label{fig:ClassicalPhaseDiagram}
\end{figure}
\end{centering}
\end{widetext}

\subsection{Quantum fluctuations}

Quantum fluctuations can {\it a priori} modify this phase diagram in two %three
main ways. First of all, if the degeneracy of the classical ground
states is accidental (that is, not related to symmetry), they can
select some of these states. This is indeed the case in the columnar
phase, where the columnar states with domain walls of only the first type
are selected already at the level of harmonic fluctuations.

\begin{figure}[bthp]
 \centering
 \includegraphics[width=8.5cm]{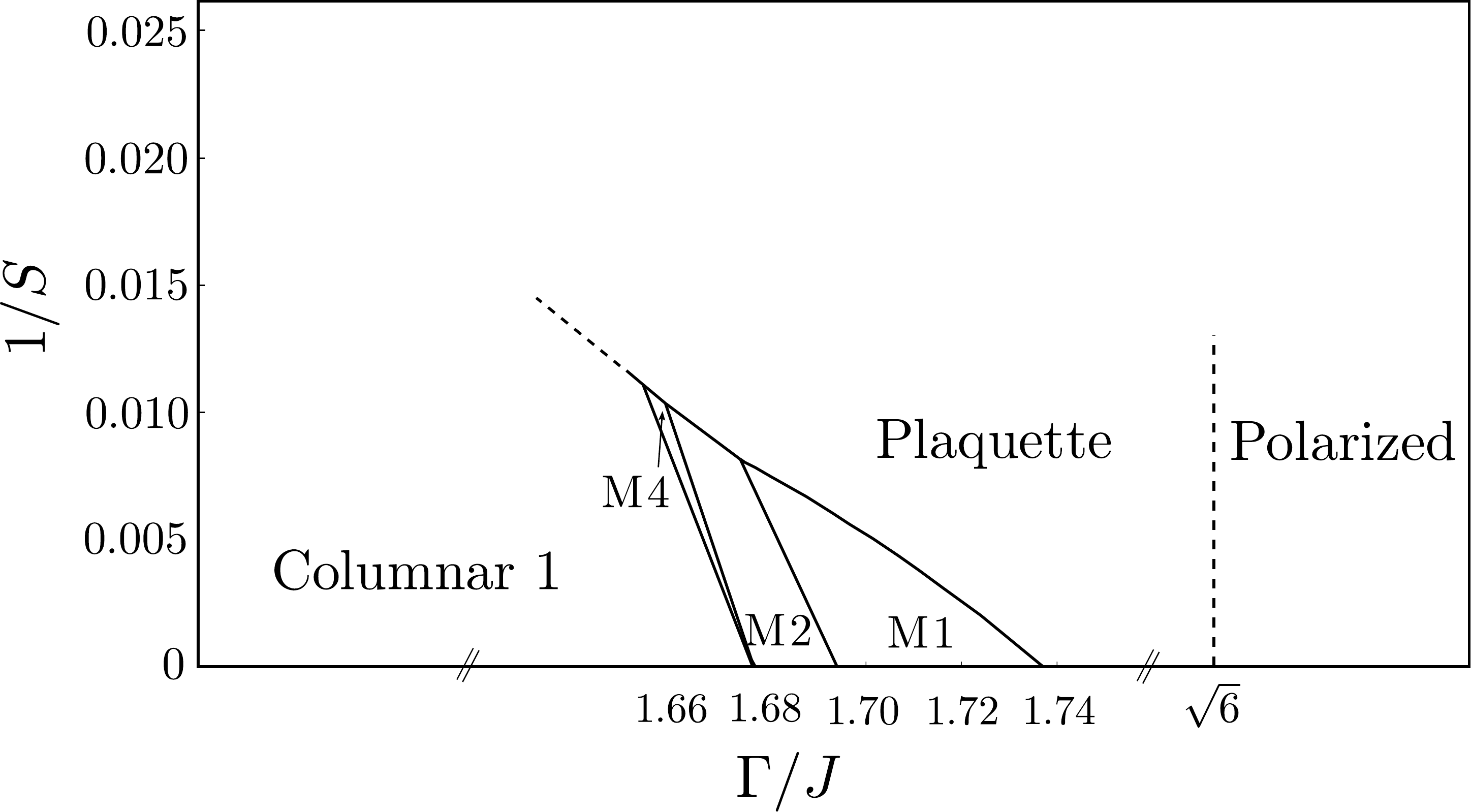} \\
 \caption{ Semiclassical corrections to the phase diagram, M$n$ $n\in\{1,2,4\}$ denote $1^{\textrm{st}}$, $2^{\textrm{nd}}$ and $4^{\textrm{th}}$ mixed states,
 zoommed for values of the field close to the accumulation point of mixed states.}
 \label{fig:SC phase diagram}
 \end{figure}

Secondly, quantum fluctuations can shift the phase boundaries.
When one takes into account only the harmonic fluctuations, this applies
only to first-order transitions. Indeed, at a first-order transition, the classical
energy is the same for the two competing configurations, but the spectra of harmonic
fluctuations are different, and one phase will in general be stabilized over the other
by zero point fluctuations. A convenient way to keep track of the stability of the
various phases with respect to quantum fluctuations is to draw a phase diagram
in the ($\Gamma/J$,$1/S$) plane (see Fig. \ref{fig:SC phase diagram}) showing
which phase has the lowest total energy.

The resulting phase diagram can be quite involved when there are many phases
in competition, and this is clearly the case here since, for $1/S = 0$, there exists an infinite sequence
of mixed phases.
%  because, if an intermediate phase disappears upon increasing $1/S$,
% one has to compare the energies of its neighboring phases to continue the phase diagram.
However, it turns out that for $1/S$ above $10^{-3}$ only three of them survive,
as is shown in Fig. \ref{fig:SC phase diagram}. All other mixed phases
exist only for $1/S\lesssim10^{-4}$
in a very narrow range of transverse magnetic field of width $\lesssim 10^{-4}J$. They
are thus invisible on the scale of Fig. \ref{fig:SC phase diagram}, which has been
adjusted to properly describe the competition between the two main phases (plaquette
and columnar). On that scale, the phase diagram consists of six phases: the polarized
phase, the plaquette phase, the first, second and fourth mixed states, and the
columnar phase. The general trend is that the plaquette phase is stabilized by
quantum fluctuations over the mixed phases as well as the columnar phase.

Note that
the transition between the plaquette and the columnar phases cannot be followed
below $\Gamma/J=1.64$ at this level of approximation because the plaquette phase
is no longer locally stable with respect to harmonic fluctuations.
The continuation of this boundary by a dashed line
in Fig.~\ref{fig:SC phase diagram} is just a guide to the eye.
To follow this line further would require to go beyond the harmonic approximation.
The transition between the plaquette and polarized phases being of the second order,
the boundary has to start vertically since, at the transition, both states
have the same quantum corrections in the harmonic approximation.
This is indicated by a vertical dashed line in Fig. \ref{fig:SC phase diagram}.
To find the curvature of this line would
require to go beyond the harmonic approximation.

% Generally critical fields {\it decrease} with
% decreasing $S$ (see Fig.\ref{fig:SC phase diagram}).}
% This results in critical fields that {\it decrease} with
% decreasing $S$ (see Fig.\ref{fig:SC phase diagram}).
% In fact, the effect is very strong, and it is impossible to follow these lines
%very far because the critical values of $\Gamma$ very quickly hit the limits of
%local stability of the %corresponding
%plaquette phase. % (see upper panel of Fig.~\ref{fig:SC phase diagram}).
%To follow these lines further would require to go beyond the harmonic approximation.

% \begin{figure}[bthp]
% \centering
%  \subfigure {\label{fig:Semi Classical Phase Diagram} \includegraphics[width=8cm]{SemiClassicalPhaseDiagram}} \\
% \vspace{0.2cm} \centering
%  \subfigure {\label{fig:Extrapolation Semi Classical Phase Diagram} \includegraphics[width=7.8cm]{ExtrapolationSemiClassicalPhaseDiagram}}
% \caption{ {(Color online)} (a) Semiclassical corrections to the phase
% diagram.
%   (b) Linear extrapolation of the semiclassical phase boundaries to $\Gamma=0$}
% \label{fig:SC phase diagram}
% \end{figure}

In view of the very strong modification of these phase boundaries upon
decreasing $S$, it is legitimate to wonder about the fate of the columnar
and mixed phases for $S=1/2$, for which the model can be mapped onto the
QDM in the limit $\Gamma/J\rightarrow 0$.
The results presented above suggest that the mixed phases
have absolutely no chance to extend to $S=\frac{1}{2}$.

Regarding the competition between the columnar and the plaquette
phases, we can get an estimate of the critical value of the spin at
which the boundary between them crosses the axis $\Gamma=0$ by looking
at the linear $1/S$ corrections starting from the point
where the two phases have the same classical energy, a point that
does not appear on the phase diagram of Fig.~\ref{fig:SC phase diagram} since 
it lies inside the $1^{\textrm{st}}$ mixed phase.
This leads to the conclusion that the columnar phase disappears above $1/S\approx 0.67$,
i.e. below $S\approx 1.49$.
Note that this should probably be considered as a lower bound in terms of $S$
since the boundary is slightly concave.
So, for $S=1/2$, the semiclassical calculation at the harmonic level
predicts only two phases: a plaquette phase up to $\Gamma/J=\sqrt{6}$,
and a polarized phase above.
The fact that we find the point $\Gamma/J=0$ to be in the region of
stability of the plaquette phase is in good agreement with the QDM, which
has been found by QMC to be in the $\sqrt{12}\times\sqrt{12}$ phase at
$V/t=0$.\cite{moessner2,ralko1}

%*************************************************************************

\section{Conclusions}

In conclusion, we have investigated the classical phase diagram of the
FFTFIM on the honeycomb lattice and how it is modified by the
semiclassical corrections induced by harmonic fluctuations. As compared to
what has been already known about the model, namely that the paramagnetic
phase is unstable at $\Gamma/J=\sqrt{6}$ towards a crystalline phase with
a large unit cell, the classical phase diagram turns out to be
surprisingly rich, with a multitude of additional phases:
a columnar phase at small transverse field and an infinite cascade of phases
of mixed columnar and plaquette character.
The phase towards which the paramagnetic phase is
unstable at $\Gamma/J=\sqrt{6}$ has been found to have the same symmetry
and periodicity as the state proposed in Ref.~\onlinecite{misguich}, but a
different structure. Both are characterized by a 24-site unit cell in the
spin language, and by a 12-site cell on the dual lattice in the dimer
language, but the state we have found has a plaquette structure. At the
classical level, the columnar phase is fully degenerate, all columnar
states having rigorously the same classical energy.

Quantum fluctuations have been found to modify this phase diagram in two
important respects. First of all, harmonic fluctuations have been shown to
partially lift the degeneracy of the columnar phase in favor of the
columnar states with only one type of domain walls. Since the
remaining degeneracy is not related to a symmetry of the model, anharmonic
corrections are expected to lift further this degeneracy. Secondly, they
modify strongly the phase boundaries, and for the ultra quantum limit,
$S=1/2$, they predict that the plaquette phase survives down to
$\Gamma\rightarrow 0$.

Going back to the original motivation of this investigation, namely the
properties of the QDM on the triangular lattice, these results deserve a
number of comments. First of all, our semiclassical approximation predicts
that the phase which is the analog of the $\sqrt{12}\times\sqrt{12}$ phase
of the QDM  has a 4-site plaquette structure. This reopens the issue of
the nature of the $\sqrt{12}\times\sqrt{12}$ phase of the QDM. According
to the results of GFQMC simulations,\cite{ralko2} possible
structures are constrained by a quasi-extinction of the dimer density
correlation function at the corner of the Brillouin zone. This has been
shown to be consistent with a uniform distribution of dimer density
inside the interior part of the 12-site hexagonal unit cell, a conclusion
somehow supported by the conclusions of Ref.~\onlinecite{misguich}
regarding the nature of the phase close to the paramagnetic phase. Now
that we know that this phase is in fact a plaquette phase, it would be
interesting to revisit the GFQMC results to see to which extent a
plaquette phase of this type might be consistent with the quasi-extinction
at the zone corner.

It is also inspiring that a columnar phase appears in the classical
solution of the FFTFIM since a similar phase is present in the QDM for
attractive interactions between dimers. We did not manage to find a
convincing connection between large $S$ in the FFTFIM and negative $V$ in
the QDM, but since we found intermediate phases between the
columnar phase and the plaquette phase in the FFTFIM, it is tempting to
speculate that such phases may also exist in the QDM.

%**************************************************************************
%**************************************************************************
\section*{Acknowledgments}

The authors acknowledge useful discussions with G. Misguich and the
financial support of the Swiss National Fund and of MaNEP.

%**************************************************************************
%**************************************************************************
\appendix
\section{Comparison of columnar and staggered states
         \label{appendix:staggered}}

Columnar states are the states which maximize $N_d$, the number of pairs
of frustrated bonds situated at the smallest possible distance from each
other [as shown in Fig. \ref{fig:2sitesDoublyFrustrated}]. In this
Appendix we want to compare the classical energy of these states with the
energy of the states in which $N_d$ is minimal, that is, is equal to zero.
In terms of dimer models such states are usually called staggered or
nonflippable states, \cite{moessner2} because they do not contain
flippable pairs of dimers.

Since in a staggered state all frustrated sites have identical
environments (with exactly one frustrated neighbor) and all non-frustrated
sites also have identical environments (with exactly two frustrated
neighbors), such a state can be described by the same two variables
$\theta_1$ and $\theta_2$ introduced in Sec. \ref{sec:classical columnar
states} for the description of a columnar state. In terms of $\theta_1$
and $\theta_2$ the energy of a staggered state can be written as
\begin{eqnarray}\label{eq:Energy Staggered States}
 E_{\rm st}& = & -\frac{N_\textrm{}}{2}\left[\frac{J}{2}\left(\cos^2{\theta_1}
 +4\cos{\theta_1}\cos{\theta_2}-\cos^2{\theta_2}\right)+\right. \nonumber\\ & &
 \hspace{20mm}\left. +\Gamma(\sin{\theta_1}+\sin{\theta_2}) \right]\;.
\end{eqnarray}
Even without minimizing $E_{\rm st}$ with respect to $\theta_1$ and
$\theta_2$ one can notice that for any $\theta_1$ and $\theta_2$
\begin{equation}
E_{\rm st}(\theta_1,\theta_2)-E_{\rm col}
(\theta_1,\theta_2)=(JN/4)(\cos\theta_1-\cos\theta_2)^2\geq
0\end{equation} and therefore
% the inequality \makebox{$E_{\rm st}
% (\theta_1,\theta_2)> E_{\rm col}(\theta_1,\theta_2)$} is satisfied for any
% $J>0$ as soon as $\cos\theta_1\neq\cos\theta_2$ and therefore
the energy of a staggered state [the minimum of $E_{\rm
st}(\theta_1,\theta_2)$] has to be higher than the energy of a columnar
state [the minimum of $E_{\rm col}(\theta_1,\theta_2)$ achieved when
$\cos\theta_1\neq\cos\theta_2$]. This proves that the maximization of
$N_d$ is always a better strategy than its minimization, even when the
ratio $\Gamma/ J$ is not small.

\end{document}